\documentclass[journal]{IEEEtran}
\usepackage{array}
\usepackage{url}
\usepackage{amsmath}
\usepackage{amsthm}
\usepackage{amssymb}
\usepackage{graphicx}
\usepackage{cite}

\makeatletter

\newcommand{\noun}[1]{\textsc{#1}}
\providecommand{\tabularnewline}{\\}

\theoremstyle{plain}
\newtheorem{thm}{\protect\theoremname}
\theoremstyle{plain}
\newtheorem{prop}[thm]{\protect\propositionname}

\@ifundefined{showcaptionsetup}{}{%
 \PassOptionsToPackage{caption=false}{subfig}}
\usepackage{subfig}
\makeatother
\providecommand{\propositionname}{Proposition}
\providecommand{\theoremname}{Theorem}

\begin{document}

\title{Privacy Management and Optimal Pricing\\in People-Centric Sensing }

\author{Mohammad~Abu~Alsheikh,~\IEEEmembership{Student~Member,~IEEE,} Dusit~Niyato,~\IEEEmembership{Fellow,~IEEE,} Derek~Leong,\\ Ping~Wang,~\IEEEmembership{Senior~Member,~IEEE,} and Zhu~Han,~\IEEEmembership{Fellow,~IEEE}
\thanks{Manuscript received September 22, 2016; revised January 12, 2017; accepted January 26, 2017.}
\thanks{M.~Abu~Alsheikh, D.~Niyato, and P.~Wang are with the School of Computer Science and Engineering, Nanyang Technological University, Singapore 639798 (emails: mohammad027@e.ntu.edu.sg, dniyato@ntu.edu.sg, and wangping@ntu.edu.sg). D.~Leong is with the Institute for Infocomm Research, Singapore 138632 (email: dleong@i2r.a-star.edu.sg). Z.~Han is with the School of Electrical and Computer Engineering, University of Houston, TX, USA 77004 (email: zhan2@uh.edu).}
}
\maketitle
\begin{abstract}
With the emerging sensing technologies such as mobile crowdsensing and Internet of Things (IoT), people-centric data can be efficiently collected and used for analytics and optimization purposes. This data is typically required to develop and render people-centric services. In this paper, we address the privacy implication, optimal pricing, and bundling of people-centric services. We first define the inverse correlation between the service quality and privacy level from data analytics perspectives. We then present the profit maximization models of selling standalone, complementary, and substitute services. Specifically, the closed-form solutions of the optimal privacy level and subscription fee are derived to maximize the gross profit of service providers. For interrelated people-centric services, we show that cooperation by service bundling of complementary services is profitable compared to the separate sales but detrimental for substitutes. We also show that the market value of a service bundle is correlated with the degree of contingency between the interrelated services. Finally, we incorporate the profit sharing models from game theory for dividing the bundling profit among the cooperative service providers.
\end{abstract}


\section{Introduction}

People-centric sensing incorporates mobile crowdsensing and the Internet of Things (IoT) to provide a platform for people to share ideas, surrounding events, and other sensing data. The collected data is required in creating and updating people-centric services\footnote{Examples of people-centric services include Waze~(\url{https://www.waze.com}) for online traffic monitoring and PatientsLikeMe~(\url{https://www.patientslikeme.com}) for healthcare experience sharing on treatments and symptoms.} offered to customers over the Internet. However, people-centric data comes with privacy threats which impede crowdsensing participants from providing their true data. A recent survey~\cite{privacy2015gigya} revealed that $90\%$ of people are concerned about their data privacy and $33\%$ are ``unsure'' whether to login with their actual identities in IoT devices. Therefore, privacy-awareness and pricing models are required in people-centric services to attain the maximum profit for service providers by jointly optimizing the privacy level and subscription fee.

People-centric services can be sold separately or together as a service bundle. Specifically, people-centric services can be interrelated with a degree of contingency. There is a joint demand for \emph{complementary services} as both services are jointly required by the customers, e.g., sentiment analysis and activity tracking. On the other hand, \emph{substitute services} are comparable in their functionality, e.g., sentiment analysis using two data analytics algorithms. To attract customers to buy the service bundle, the bundle is offered at a lower subscription fee compared to the sum of subscription fees of the separately sold services. In this paper, we examine three important questions that arise regarding pricing and bundling of people-centric services. Firstly, what are the optimal subscription fee and privacy level required for profit maximization? Secondly, is service bundling effective for profit maximization in people-centric services and does it produce more gross profit compared to the separate sales? Thirdly, how is the bundling profit divided among the interrelated services?

This paper provides a framework for optimal pricing and privacy management in people-centric services. We assume that service providers are rational, refuse to offer their services unless they can recover the total data cost, and seek to maximize their own gross profits from selling people-centric services. The key contributions of this paper can be summarized as follows:
\begin{itemize}
\item We first present a model to define the utility of data in people-centric
sensing. Using real-world datasets, we show that this utility model
captures the inverse correlation between the privacy level and service
quality where data analytics is extensively utilized.
\item We formulate a pricing scheme and profit maximization model for separately selling privacy-aware people-centric services. The data is collected from crowdsensing participants and used in training and updating people-centric services which are offered to customers for a subscription fee. The optimal subscription fee and privacy level are optimized to maximize the gross profit of service providers.
\item We then propose a bundling scheme for virtually packaging privacy-aware people-centric services as complements or substitutes. The subscription fee of the service bundle and the privacy levels of the two services are optimized by a profit maximization model. We introduce a profit allocation model for sharing the profit resulting from the service bundle among the individual bundled services. The efficient and fair profit allocations encourage collaboration among service providers in providing service bundles of complementary services. Nonetheless, bundling of substitute services is detrimental and yields low profit allocations.
\end{itemize}

The rest of this paper is structured as follows. Section~\ref{sec:related_work} reviews the related work. Then, the system model and a brief overview of people-centric sensing are presented in Section~\ref{sec:system_model}. Section~\ref{sec:seperate_sales} introduces the optimization model of separately selling privacy-aware people-centric services. Next, the profit maximization models of selling bundled complementary and substitute services with privacy awareness are presented in Section~\ref{sec:bundling_sales}. Section~\ref{sec:experimental_results} presents and analyzes the experimental results. The paper is finally concluded in Section~\ref{sec:conclusions}.

\section{Related Work\label{sec:related_work}}

People are the primary focus in people-centric sensing, which has applications in transportation systems~\cite{zhang2011data}, assistive healthcare~\cite{giannetsos2011people}, and urban monitoring\cite{zhang2013verifiable}, just to name a few. In this section, we first review related work on pricing models in sensing and communication systems. Then, we discuss the crucial issue of privacy awareness in people-centric sensing. Finally, we review related work on pricing and incentive mechanisms for mobile crowdsensing.

\subsection{Pricing in Sensing and Communication Systems}

Pricing models ensure financial stability and resiliency in sensing and communication systems. The authors in~\cite{he2006pricing} presented a cooperative pricing model for Internet providers offering the Internet service under one coalition. The cooperative pricing increases the profit and encourages the Internet providers to upgrade their network connections. A pricing scheme based on the customer data usage of Internet services was introduced in~\cite{mausage2016}. Unlike flat-rate pricing, the usage-based pricing enables a fair allocation of the Internet resource among the customers. The authors in~\cite{duan2013economics} presented a pricing model of accessing femtocell and macrocell by mobile devices which enables high service quality and maximizes the profit of network operators. The authors in~\cite{huang2010optimality} proposed a pricing and transmission scheduling models to maximize the profit of accessing a wireless network by mobile customers. The customer demand is modeled as a Markov chain where applying only two price options is found sufficient for each demand state.

The pricing models of people-centric services is more challenging compared to other communication systems. Specifically, the resources and utility of people-centric services are not easily measured as other systems, e.g., the bandwidth and connection speed are easily defined for an Internet service.

\subsection{Privacy-Awareness in People-Centric Sensing}

People-centric sensing incorporates the IoT and mobile crowdsensing and considers the privacy of people. In~\cite{giannetsos2011people}, the authors pointed out that the data privacy is a key challenge for people-centric sensing in assistive healthcare. The data traces in healthcare systems typically include personal habits and traits of users and introduce the risk of privacy disclosure. The authors in~\cite{zhang2013verifiable} discussed people-centric sensing as a data collection method in urban areas. The crowdsensing participants apply additive and non-additive aggregation models, such as averaging, to ensure that their true data cannot be disclosed. In~\cite{cornelius2008anonysense}, anonymous data collection in people-centric sensing was presented as a privacy preserving model. The privacy model is applied with low resource demand of CPU and network bandwidth as mobile devices are resource-constrained. Additionally, there are other works related to location privacy in people-centric sensing, e.g.,~\cite{krumm2007inference,shin2011anonysense,pournajaf2014spatial}, which prevent localization attacks and protect the privacy of participants in spatial tasks. Spatial cloaking, adding noise, and rounding are commonly used in the literature for location obfuscation.

There are a few models to define the data privacy including $k$-anonymity~\cite{sweeney2002k}, $l$-diversity~\cite{machanavajjhala2007diversity}, and differential privacy~\cite{dwork2008differential}. The k-anonymity model requires that a person cannot be identified from at least a group of $k-1$ other people. However, $k$-anonymity does not guarantee privacy against background knowledge attacks, e.g., merging quasi-sensitive attributes with other datasets. The $l$-diversity model~\cite{machanavajjhala2007diversity} addresses background knowledge attacks and ensures that any quasi-sensitive group has at least $l$ ``well-represented'' values. Differential privacy~\cite{dwork2008differential} does not include assumptions on the adversary's background knowledge. Instead, differential privacy requires that the probability distribution of shared data from a privacy preserving method does not change by adding one person's data. This ensures that the adversary cannot recognize a particular person from differentially-private data. Random noise is generally added to the raw data to meet the definition of differential privacy~\cite{mohammed2011differentially,xiao2011differential,abadi2016deep}.

\subsection{Pricing and Incentive Mechanisms for Mobile Crowdsensing }

Pricing and incentive mechanisms are required to encourage participation in data collection. A reward-based incentive mechanism for mobile crowdsensing was presented in~\cite{duan2012incentive}. The reservation wages of the participants are utilized to reduce the total data cost by selecting the sufficient set of participants with the lowest rates. In~\cite{wen2015quality}, the participants are paid according to their reliability. The reliability is defined as a probabilistic process and measured based on the historical records of the participants in completing crowdsensing tasks. The authors in~\cite{luo2015incentive} considered the heterogeneity of crowdsensing participants and proposed asymmetric payment model which encourages competition among the participants. In~\cite{niyato2016market}, the authors introduced a profit maximization and pricing model to optimize the amount of data that should be bought from the sensing participants.

None of the existing papers on people-centric sensing in the literature consider the problem of jointly optimizing the pricing and privacy level in people-centric services where data analytics is heavily applied. Moreover, existing works do not consider bundling interrelated people-centric services as complements or substitutes. Therefore, there is a practical demand for privacy-aware pricing, bundling, and profit allocation models which are the major contributions of this paper.

\section{People-Centric Big Data: System Model\label{sec:system_model}}

In this section, we first discuss the system model of people-centric sensing and services considered in this paper. We then briefly present the market strategy of bundling people-centric services. Finally, we introduce the tradeoff between the privacy level and service quality in data analytics perspectives.

\subsection{People-Centric Services}

\begin{figure*}
\begin{centering}
\includegraphics[width=0.9\linewidth,trim=0cm 0.5cm 0cm 0cm]{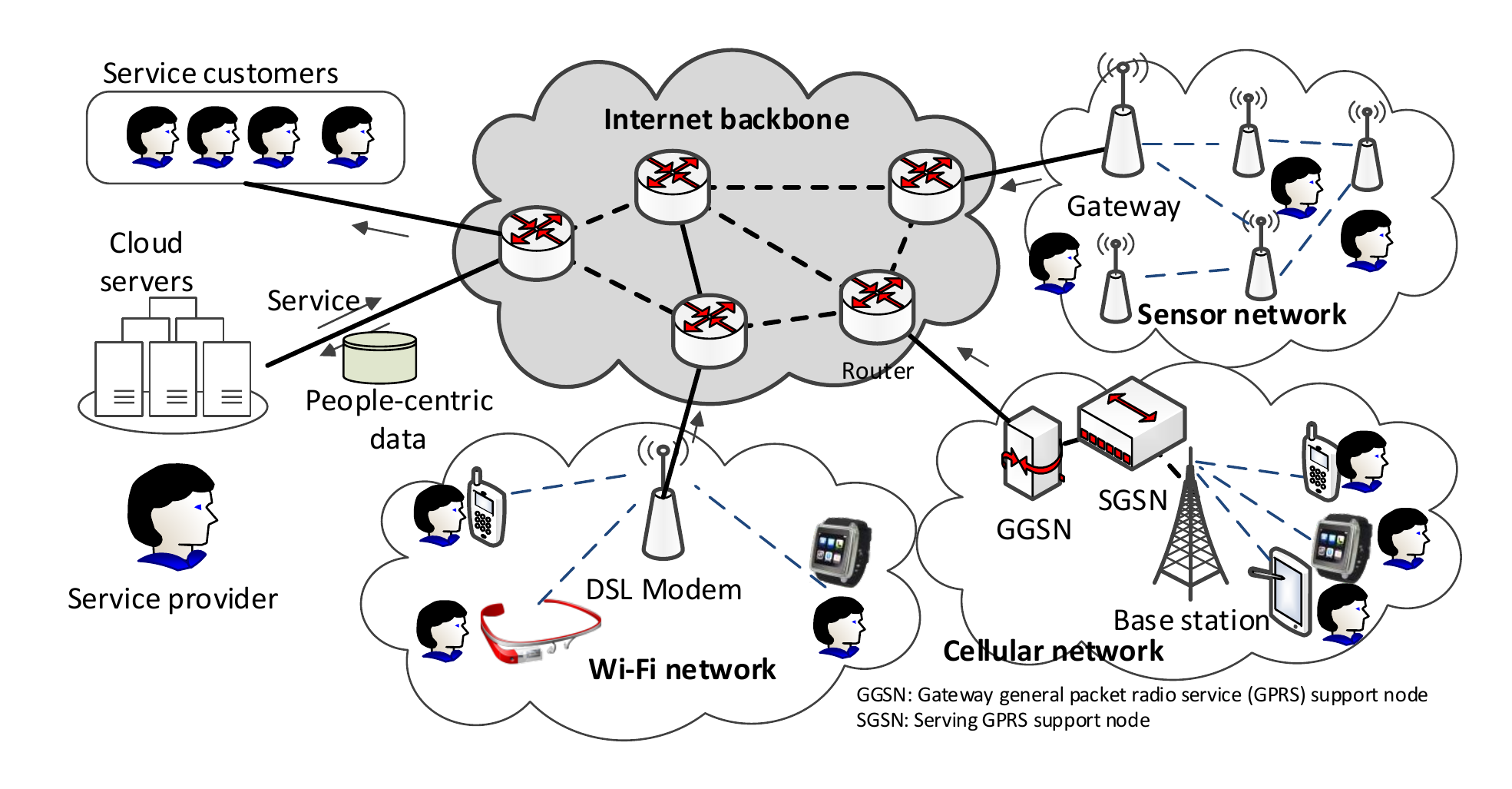}
\par\end{centering}
\caption{System model of people-centric data collection and service development.\label{fig:people_centric_sensing}}
\end{figure*}

Figure~\ref{fig:people_centric_sensing} shows the system model of people-centric sensing and services. People share their crowdsensing data through a massive system of mobile devices and IoT gadgets. In particular, the proliferation of sensors, e.g., cameras, microphones, and accelerometers, in mobile devices enables people to participate in cooperative sensing of events and phenomena. Besides, people intelligence can be incorporated in the sensing process which helps in collecting complex and rich data. Other data regarding people can also come from conventional sensor networks. The network connections in people-centric sensing include various technologies such as cellular and Wi-Fi networks. In the following we describe the major entities in people-centric services under consideration:
\begin{itemize}
\item \emph{Crowdsensing participants} are the source of the data. We consider a people-centric service with $N$ crowdsensing participants where each participant $i$ produces privacy preserving data denoted as follows:
\begin{equation}
\begin{array}{cc}
y_{i}=x_{i}+z_{i}, & i=1,\ldots,N\end{array}
\end{equation}
where $y_{i}$ is the noisy data, $x_{i}$ is the true data, and $z_{i}$ is the added noise component. We assume that the noise components $\left\{ z_{i}\right\} _{i=1}^{N}$ are independent Gaussian random components with zero mean and a variance of $\sigma_{z}^{2}$. We also denote this data in the vector form as $\mathbf{y}\in\mathbb{R}^{N}$, $\mathbf{x}\in\mathbb{R}^{N}$, and $\mathbf{z}\in\mathbb{R}^{N}$ such that $\mathbf{z}\sim\mathbb{N}\left(0,\sigma_{z}^{2}\mathbb{I}_{N}\right)$ where $\mathbb{I}_{N}$ is the identity matrix of size $N$. To attain high service quality, the participant stochastically sends its true data $\mathbf{x}$ to the people-centric service with a probability of $\mathbb{P}\left(\text{true data}\right)$. For the remainder of this paper, we define the privacy level $r$ to be equal to the probability of sending the noisy data $\mathbf{y}$ instead of the true data $\mathbf{x}$ such that $\mathbb{P}\left(\text{true data}\right)=1-r$. Our system model is general and can incorporate any privacy definition such as $k$-anonymity~\cite{sweeney2002k}, $l$-diversity~\cite{machanavajjhala2007diversity}, and differential privacy~\cite{dwork2008differential}. Each participant, whether it is a member of the public or data warehouses, has a \emph{reservation wage} $c$ which is the lowest payment required to work as a data collector.
\item A\emph{ service provider} buys people-centric data from the crowdsensing participants and applies data analytics to build the people-centric service. This service is hosted at one of the cloud computing platforms such as Microsoft Azure and Amazon Web Services (AWS). To cover the operation cost, the service provider charges a ``subscription fee'' $p_{s}$ to customers who access the people-centric service. Moreover, the service provider decides the ``privacy level'' $r$ at which the data should be collected by the crowdsensing participants. For gross profit maximization, the service provider jointly optimizes the subscription fee and privacy level of its service.
\item \emph{Customers} are the users of the people-centric service. Each customer has a \emph{reservation price} $\theta$ which is the maximum price at which that particular customer will buy the people-centric service. A customer considers both the reservation price $\theta$, service quality $u$, and subscription fee $p_{s}$ when making its buying decision. In particular, a customer buys the service if the inequality $\theta\geq\frac{p_{s}}{u}$ holds.
\end{itemize}
As summarized in Table~\ref{tab:comparision_people_conventional}, this people-centric sensing model overcomes the limitations of conventional sensing systems based on sensor networks only. However, people-centric sensing comes with the privacy challenge which should be considered in optimal pricing and profit maximization.

\begin{table*}
\begin{centering}
\caption{Comparison of people-centric sensing and conventional sensing.\label{tab:comparision_people_conventional}}
\begin{tabular}{|l|>{\raggedright}p{5cm}|>{\raggedright}p{5cm}|}
\hline 
\textbf{\noun{Aspect}} & \textbf{\noun{people-centric sensing}} & \textbf{\noun{conventional sensing}}\tabularnewline
\hline 
\hline 
{\small{}Deployment} & {\small{}Mobile devices owned by participants} & {\small{}Sensor nodes typically owned by service providers}\tabularnewline
\hline 
{\small{}People engagement} & {\small{}Human in the loop} & {\small{}Machines only}\tabularnewline
\hline 
{\small{}Mobility} & {\small{}Move with people} & {\small{}Static or limited mobility}\tabularnewline
\hline 
{\small{}Key challenge} & {\small{}Data privacy} & {\small{}Energy conservation}\tabularnewline
\hline 
\end{tabular}
\par\end{centering}

\end{table*}

\subsection{Bundling Interrelated Services}

\begin{figure}
\begin{centering}
\includegraphics[width=1\columnwidth,trim=0cm 0.5cm 1cm 0cm]{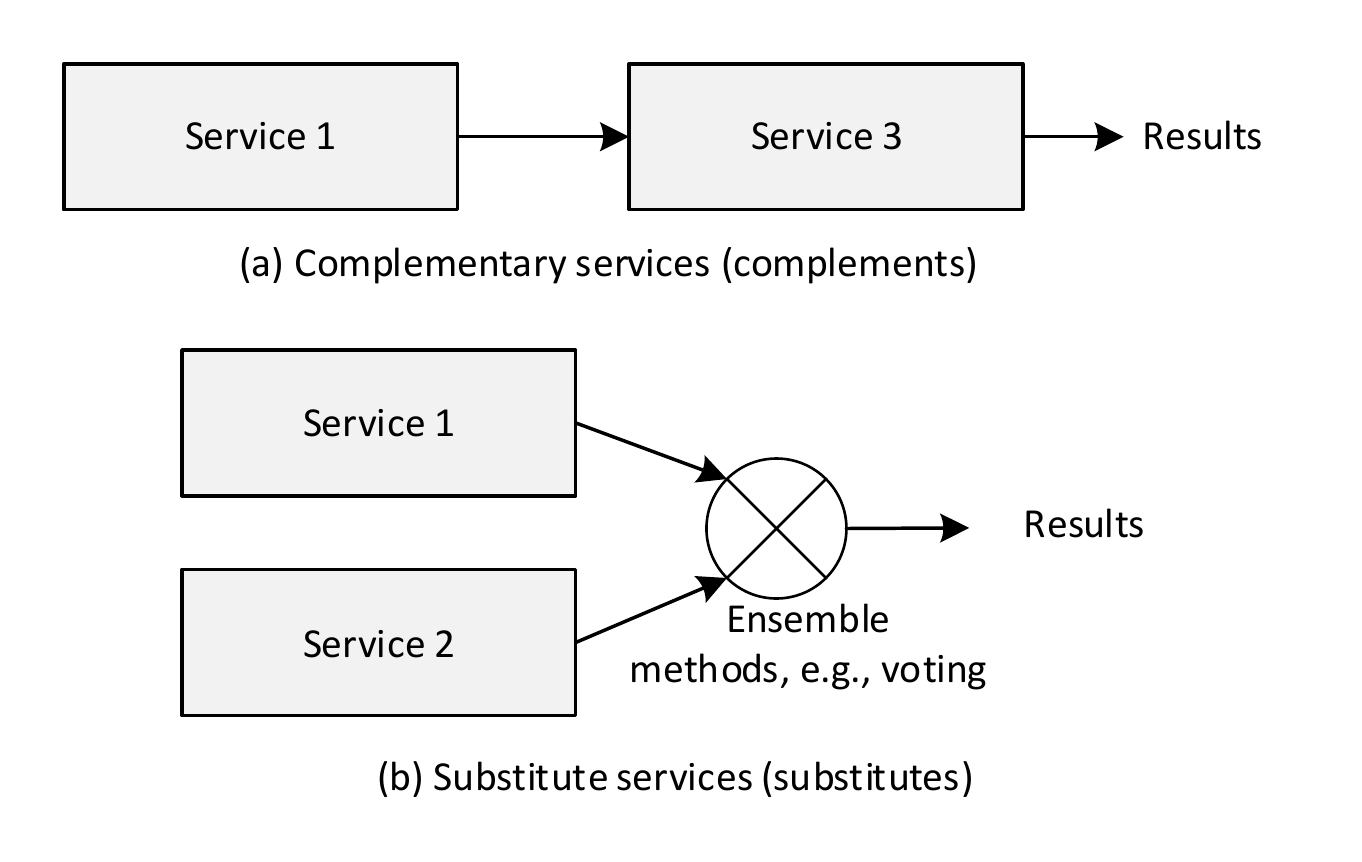}
\par\end{centering}
\caption{Interrelated services as complements and substitutes.\label{fig:interrelated_services}}
\end{figure}

People-centric services can be interrelated and sold as one service bundle\footnote{Product bundling is an effective marketing strategy of selling products in one package, e.g., Microsoft Office includes Microsoft Word, Excel, and PowerPoint. }. Figure~\ref{fig:interrelated_services} illustrates the interrelation among people-centric services from the perspectives of customers. Complementary services are associated and concurrently required to achieve the objectives of customers. For example, both sentiment analysis~\cite{go2009twitter} and activity tracking~\cite{kwapisz2011activity} are typically required to provide in-depth understanding of human-intense mobile systems. On the other hand, substitute services have similar or comparable functionalities which decrease the customer's willingness in buying both services. If a customer buys one of the substitute services, that customer will probably not buy its paired service. For example, sentiment analysis using the data analytics models of deep learning~\cite{Goodfellow2016deep} and random forests~\cite{breiman2001random} are substitute. In some scenarios, the customers buy both substitute services to improve the performance of service quality, e.g., a mixture of models in ensemble learning~\cite{polikar2006ensemble}.

\subsection{The Quality-Privacy Tradeoff}

There are several reasons for examining the privacy and optimal pricing of people-centric services. Firstly, privacy is a common concern of people. Secondly, reservation wages of crowdsensing participants are correlated with the utility of data and service quality. The quality of data analytics is inversely proportional to privacy level~\cite{aggarwal2008general}. Thirdly, the customers infer both the service quality and subscription fee when deciding whether to buy a people-centric service. The utility function of data $u(\cdot)$ in people-centric services should meet the following empirical assumptions:
\begin{itemize}
\item $u(\cdot)$ is nonnegative. This is rational as the service quality cannot be negative.
\item $u(\cdot)$ is inversely proportional to the privacy level $r\in[0,1]$ such that $\frac{\partial u(\cdot)}{\partial r}<0$. This empirical assumption is required as increasing the privacy level decreases the quality of data analytics~\cite{aggarwal2008general}.
\item $u(\cdot)$ is convex and decreases at an increasing rate over the privacy level such that $\frac{\partial^{2}u(\cdot)}{\partial r^{2}}<0$. This assumption reflects the empirical change of service quality at varying privacy levels.
\end{itemize}
Based on these empirical assumptions and to facilitate our optimization modeling, we propose the following utility function:
\begin{equation}
u(r;\mathbf{\alpha})=\alpha_{1}-\alpha_{2}\exp\left(\alpha_{3}r\right),\label{eq:accuracy-privacy_fun}
\end{equation}
where $r$ is the privacy level. $\alpha_{1}$, $\alpha_{2}$, and $\alpha_{3}$ are the curve fitting parameters of the utility function to real-world experiments, i.e., the ground truth. Big data platforms, e.g., Apache Mahout~\cite{mahout2015scalable} and MLlib~\cite{meng2015mllib}, can be used for running the real-world experiments at scale. In particular, a set of $B$ real-world experiments $\left\{ \left(r^{(i)},\tau^{(i)}\right)\right\} _{i=1}^{B}$ are executed at varying privacy level $r^{(i)}$ resulting in the real-world service quality of $\tau^{(i)}$, where $r^{(i+1)}>r^{(i)}\geq0$. $\alpha_{1}$, $\alpha_{2}$, and $\alpha_{3}$ are obtained by minimizing the residuals of a nonlinear least squares fitting as follows:
\begin{equation}
\underset{\alpha}{\text{minimize }}\sum_{i=1}^{B}\left\Vert u(r^{(i)};\mathbf{\alpha})-\tau^{(i)}\right\Vert ^{2}.\label{eq:nonlinear_least_squares}
\end{equation}
(\ref{eq:nonlinear_least_squares}) can be solved iteratively to find the best fitting parameters $\alpha$~\cite{strutz2010data}. We denote $u(r;\mathbf{\alpha})$ as $u$ whenever it does not cause confusion. In Section~\ref{sub:accuracy_privacy_tradeoff}, we show the validity of (\ref{eq:accuracy-privacy_fun}) in capturing the quality-privacy tradeoff of people-centric services trained on real-world datasets.

\begin{figure}
\begin{centering}
\includegraphics[width=1\linewidth,trim=0 0.5cm 0 0]{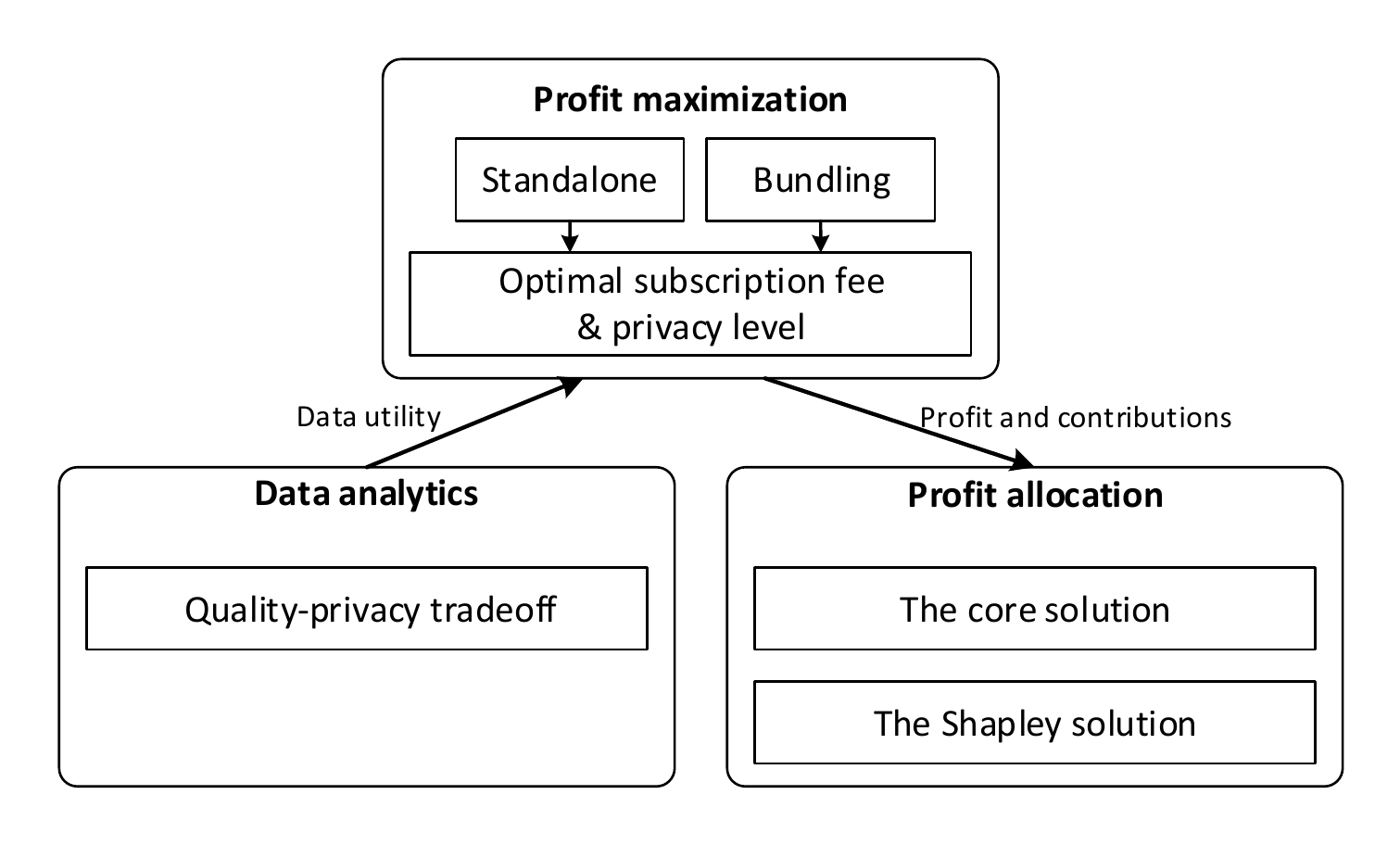}
\par\end{centering}
\caption{Components of the optimal pricing and privacy management framework for people-centric sensing.\label{fig:profit_max_framework}}
\end{figure}

Figure~\ref{fig:profit_max_framework} shows the key components of the optimal pricing and privacy management framework proposed in this paper. These components are executed iteratively. The framework is initiated by defining the data utility using the form expressed in (\ref{eq:accuracy-privacy_fun}). Then, the profit maximization models are executed to obtain the optimal subscription fee and privacy level. These profit maximization models are presented in Sections~\ref{sec:seperate_sales} and \ref{sec:bundling_sales} for separate and bundling sales, respectively. For bundling, the profit allocation models presented in Section~\ref{sub:profit_sharing} are performed. Then, the service provider decides whether the service bundling is effective to attain the maximum profit. The key notations and definitions used throughout the paper are defined in Table~\ref{tab:key_notations}.

\begin{table}
\caption{List of key symbols and notations.\label{tab:key_notations}}

\centering{}%
\begin{tabular}{|>{\raggedright}p{0.18\columnwidth}|>{\raggedright}p{0.72\columnwidth}|}
\hline 
\textbf{\noun{Notation}} & \textbf{\noun{Definition}}\tabularnewline
\hline 
\hline 
$p_{s}$ & {\small{}Subscription fee required to access a people-centric service
under separate selling}\tabularnewline
\hline 
$r$ & {\small{}Privacy level at which the data is collected from crowdsensing participants}\tabularnewline
\hline 
$F(r,p_{s})$ & {\small{}Gross profit resulting from the separate sales of a people-centric
service under $p_{s}$ and $r$}\tabularnewline
\hline 
$u(r;\alpha)$ & {\small{}Service quality with curve fitting parameters $\alpha$ and
privacy level $r$. $u_{k}$ is the service quality of the $k$-th
service in a service bundle }\tabularnewline
\hline 
$M$ & {\small{}Number of customers willing to buy a people-centric service}\tabularnewline
\hline 
$N$ & {\small{}Number of crowdsensing participants}\tabularnewline
\hline 
$c$ & {\small{}Reservation wage of one crowdsensing participant}\tabularnewline
\hline 
$\theta$ & {\small{}Reservation price defining the maximum price at which a customer
is willing to buy a people-centric service. $\theta$ has a cumulative
distribution function of $\Phi_{\theta}$}\tabularnewline
\hline 
$\mathcal{L}(\cdot)$ & {\small{}Lagrangian of the profit function $F(\cdot)$}\tabularnewline
\hline 
$D_{k}$ & {\small{}$k$-th leading principal minor of the profit function's
Hessian matrix}\tabularnewline
\hline 
$\gamma$ & {\small{}Degree of contingency of two interrelated services}\tabularnewline
\hline 
$S_{b}$ & {\small{}Service bundle}\tabularnewline
\hline 
$p_{b}$ & {\small{}Subscription fee of a service bundle $S_{b}$ containing
two complementary or substitute services}\tabularnewline
\hline 
$G_{c}(r_{1},r_{2},p_{b})$ & {\small{}Gross bundling profit of complementary services with privacy
levels $r_{1}$ and $r_{2}$ offered at a subscription fee of $p_{b}$}\tabularnewline
\hline 
$G_{s}(r_{1},r_{2},p_{b})$ & {\small{}Gross bundling profit of substitute services with privacy
levels $r_{1}$ and $r_{2}$ offered at a subscription fee of $p_{b}$}\tabularnewline
\hline 
$\varphi_{k}$ & {\small{}Payoff allocation assigned using the core solution to the
$k$-th service in $S_{b}$}\tabularnewline
\hline 
$\eta_{k}$ & {\small{}Payoff allocation assigned using the Shapley value concept
to the $k$-th service in $S_{b}$}\tabularnewline
\hline 
\end{tabular}
\end{table}

\section{Optimal Pricing in People-Centric Services\label{sec:seperate_sales}}

In this section, we first present the market model of selling people-centric service separately. Then, we introduce the profit maximization model with privacy awareness. Finally, the closed-form solutions of the subscription fee and privacy level are derived and proved to be globally optimal.

\subsection{Gross Profit Maximization}

The system model under consideration in this section is shown in Figure~\ref{fig:people_centric_sensing}. The service provider collects data from crowdsensing participants at a privacy level $r$. The data is essential to train and update a people-centric service offered to customers paying a subscription fee $p_{s}$. The gross profit $F(\cdot)$ of selling the people-centric service can be defined mathematically as follows:
\begin{equation}
F(r,p_{s})=\underbrace{Mp_{s}\mathbb{P}\left(\theta\geq\frac{p_{s}}{u}\right)}_{\text{Subscription revenue}}-\underbrace{Nc\mathbb{P}\left(\text{true data}\right)}_{\text{Total data cost}},\label{eq:profit_function_seperate_1}
\end{equation}
where $M$ is the number of potential customers, $p_{s}$ is the service subscription fee, $r$ is the privacy level, $c$ is the reservation wage of crowdsensing participants, and $N$ is the number of potential crowdsensing participants. The gross profit $F(\cdot)$ is the difference between the subscription revenue and total data cost. The operational cost of the service, such as the computing cost, is neglected. The first term of (\ref{eq:profit_function_seperate_1}) defines the subscription revenue resulting from offering the service at a subscription fee of $p_{s}$ and service quality $u$. $\mathbb{P}\left(\theta\geq\frac{p_{s}}{u}\right)=1-\Phi_{\theta}\left(\frac{p_{s}}{u}\right)$ is the probability for a customer to buy the service after inferring both $p_{s}$ and $u$ which can be calculated from the complementary cumulative distribution function. The total data cost is defined in the second term of (\ref{eq:profit_function_seperate_1}) to be proportional to the probability of sending the true data $\mathbb{P}\left(\text{true data}\right)$. This is rational as the service quality and gross profit are negatively affected by increasing the privacy level, and thus the service provider should also pay less for the noisy data. Assuming that $\theta$ follows a uniform distribution over the interval $[0,1]$, (\ref{eq:profit_function_seperate_1}) can be written as follows: 
\begin{equation}
F(r,p_{s})=Mp_{s}\left(1-\frac{p_{s}}{\alpha_{1}-\alpha_{2}\exp\left(\alpha_{3}r\right)}\right)-Nc\left(1-r\right).\label{eq:profit_function_seperate}
\end{equation}

The profit maximization problem can be formulated as follows:
\begin{equation}
\begin{aligned}\underset{r,p_{s}}{\text{maximize }} & F(r,p_{s})\\
\text{subject to } & C1:p_{s}\geq0,\\
 & C2:r\geq0.
\end{aligned}
\label{eq:seperate_sale_optimization}
\end{equation}
The objective of (\ref{eq:seperate_sale_optimization}) is to maximize the gross profit by jointly optimizing $p_{s}$ and $r$. The constraints $C1$ and $C2$ are required to ensure nonnegative solutions of $p_{s}$ and $r$, respectively. We next provide the closed-form solution $(p_{s}^{*},r^{*})$ of this profit maximization problem and prove their global optimality. 

\subsection{Optimal Subscription Fee and Privacy Level}

We apply the Karush-Kuhn-Tucker (KKT) conditions which are sufficient for primal-dual optimality of concave functions~\cite{chong2013introduction}. Based on (\ref{eq:seperate_sale_optimization}), we formulate the Lagrangian dual function as follows:
\begin{equation}
\mathcal{L}\left(r,p_{s},\lambda_{1},\lambda_{2}\right)=F(r,p_{s})+\lambda_{1}p_{s}+\lambda_{2}r,\label{eq:seperate_sale_lagrange}
\end{equation}
where $\lambda_{1}\geq0$ and $\lambda_{2}\geq0$ are the Lagrange multipliers (dual variables) associated with the constraints $C_{1}$ and $C_{2}$, respectively.
\begin{prop}
The closed-form solutions $p_{s}^{*}$ and $r^{*}$ of (\ref{eq:seperate_sale_optimization}) can be derived as follows:
\begin{eqnarray}
p_{s}^{*} & = & \frac{M\alpha_{1}\alpha_{3}-4Nc}{2M\alpha_{3}},\label{eq:p_s_optimal}\\
r^{*} & = & \frac{1}{\alpha_{3}}\log\left(\frac{4Nc}{M\alpha_{2}\alpha_{3}}\right),\label{eq:r_optimal}
\end{eqnarray}
where $\lambda_{1}=\lambda_{2}=0.$\end{prop}
\begin{IEEEproof}
To achieve this result, the first derivatives of (\ref{eq:seperate_sale_lagrange}) with respect to $p_{s}$ and $r$ are found as follows:
\begin{eqnarray}
\frac{\partial\mathcal{L}\left(\cdot\right)}{\partial p_{s}} & = & M+\lambda_{1}-\frac{2Mp_{s}}{\alpha_{1}-\alpha_{2}\exp\left(\alpha_{3}r\right)},\\
\frac{\partial\mathcal{L}\left(\cdot\right)}{\partial r} & = & Nc+\lambda_{2}-\frac{M\alpha_{2}\alpha_{3}p_{s}^{2}\exp\left(\alpha_{3}r\right)}{\left(\alpha_{1}-\alpha_{2}\exp\left(\alpha_{3}r\right)\right)^{2}}.
\end{eqnarray}
The closed-form solutions in (\ref{eq:p_s_optimal}) and (\ref{eq:r_optimal}) with inactive inequality constraints ($\lambda_{1}=\lambda_{2}=0$) can then be deduced by setting both derivatives to zero and solving the resulting system of equations.
\end{IEEEproof}
\begin{prop}
$F(r,p_{s})$ is concave. The closed-form solutions $p_{s}^{*}$ and $r^{*}$ given in (\ref{eq:p_s_optimal}) and (\ref{eq:r_optimal}), respectively, are globally optimal.
\end{prop}
\begin{IEEEproof}
We use Sylvester's criterion as a sufficient condition to show that the Hessian matrix of $F(r,p_{s})$ is negative semidefinite, and hence the concavity of $F(r,p_{s})$ can be deduced~\cite{chong2013introduction}. In particular, the Hessian matrix $\mathbf{H}_{F}$ of $F(r,p_{s})$ is found as in~(\ref{eq:seperate_hessian_matrix}), shown at the top of the next page.
\begin{figure*}[t]
\begin{equation}
\mathbf{H}_{F}=\left[\begin{array}{cc}
-\frac{2M}{\alpha_{1}-\alpha_{2}\exp\left(\alpha_{3}r\right)} & -\frac{2M\alpha_{2}\alpha_{3}p_{s}\exp\left(\alpha_{3}r\right)}{\left(\alpha_{1}-\alpha_{2}\exp\left(\alpha_{3}r\right)\right)^{2}}\\
-\frac{2M\alpha_{2}\alpha_{3}p_{s}\exp\left(\alpha_{3}r\right)}{\left(\alpha_{1}-\alpha_{2}\exp\left(\alpha_{3}r\right)\right)^{2}} & -\frac{2M\alpha_{2}^{2}\alpha_{3}^{2}p_{s}^{2}\exp\left(2\alpha_{3}r\right)}{\left(\alpha_{1}-\alpha_{2}\exp\left(\alpha_{3}r\right)\right)^{3}}-\frac{M\alpha_{2}\alpha_{3}^{2}p_{s}^{2}\exp\left(\alpha_{3}r\right)}{\left(\alpha_{1}-\alpha_{2}\exp\left(\alpha_{3}r\right)\right)^{2}}
\end{array}\right].
\label{eq:seperate_hessian_matrix}
\end{equation}
\end{figure*}
Let $D_{k}$ be the $k$-th leading principal minor of $\mathbf{H}_{F}$, where $k=1,2$. $\mathbf{H}_{F}$ is negative semidefinite if $(-1)^{k}D_{k}\geq0$~\cite{chong2013introduction}. The leading principal minors of $\mathbf{H}_{F}$ are 
\begin{eqnarray}
D_{1} & = & -\frac{2M}{\alpha_{1}-\alpha_{2}\exp\left(\alpha_{3}r\right)}\leq0,\\
D_{2} & = & \frac{2M^{2}\alpha_{2}\alpha_{3}^{2}p_{s}^{2}\exp\left(\alpha_{3}r\right)}{\left(\alpha_{1}-\alpha_{2}\exp\left(\alpha_{3}r\right)\right)^{3}}\geq0,
\end{eqnarray}
which alternate in sign with $D_{1}$ being nonpositive. Therefore, $\mathbf{H}_{F}$ is negative semidefinite and $F(r,p_{s})$ is concave. By concavity, the closed-form solutions expressed in (\ref{eq:p_s_optimal}) and (\ref{eq:r_optimal}) are globally optimal.
\end{IEEEproof}

\subsection{Special Case: Fixed Privacy Level\label{sub:seperate_fixed_privacy}}

We next discuss the special case when the service provider cannot control the privacy level, e.g., $r$ is fixed by a legislative court\footnote{The European Commission (\url{http://ec.europa.eu}), for example, regularly revises a set of regulations to protect the data privacy of citizens in the European Union.}. In such a case, the profit of the service provider can be defined as in (\ref{eq:profit_function_seperate}) with $r$ being fixed. For profit maximization, the service provider responds by selecting the optimal subscription fee as deduced in the following proposition.
\begin{prop}
When the privacy level $r$ is fixed by an external entity, the optimal subscription fee is found as follows:
\begin{equation}
p_{s}^{*}=\frac{\alpha_{1}-\alpha_{2}\exp\left(\alpha_{3}r\right)}{2},\label{eq:p_s_optimal_fixed_privacy}
\end{equation}
which is globally optimal.\end{prop}
\begin{IEEEproof}
The second derivatives of $F(r,p_{s})$ given in (\ref{eq:profit_function_seperate}) with respect to $p_{s}$ is defined as follows:
\begin{equation}
\frac{\partial^{2}F\left(\cdot\right)}{\partial p_{s}^{2}}=-\frac{2M}{\alpha_{1}-\alpha_{2}\exp\left(\alpha_{3}r\right)}\leq0,
\end{equation}
which is always nonpositive. Thus, $F(r,p_{s})$ is concave and the solution in (\ref{eq:p_s_optimal_fixed_privacy}) of the fixed privacy problem is globally optimal.
\end{IEEEproof}

\begin{figure}
\begin{centering}
\includegraphics[width=1\columnwidth,trim=1cm 1cm 1cm 0cm]{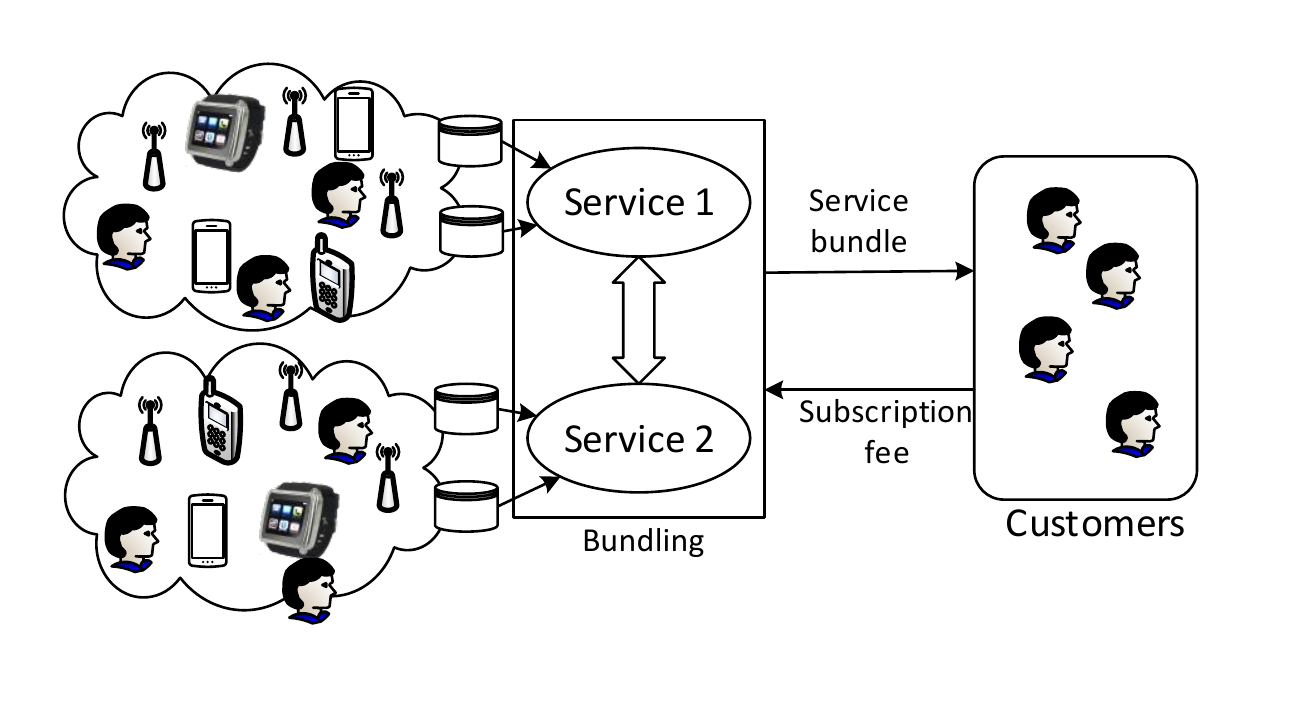}
\par\end{centering}
\caption{System model of bundled people-centric services.\label{fig:bundle_system}}
\end{figure}

\section{Interrelated People-Centric Services\label{sec:bundling_sales}}

People-centric services can be interrelated as complements and substitutes as shown in Figure~\ref{fig:interrelated_services}. The joint optimization of the subscription fee and privacy levels in a service bundle is introduced in this section. Firstly, we present the system model and define the degree of contingency in service bundling. Secondly, we present the profit maximization models and the closed-form solutions of selling service bundles as complements and substitutes, respectively. The closed-form solutions are also shown to be globally optimal. Finally, we define the profit shares that should be allocated to each service within the bundle.

\begin{figure*}
\begin{centering}
\includegraphics[width=0.7\linewidth,trim=1cm 0cm 1cm 0cm]{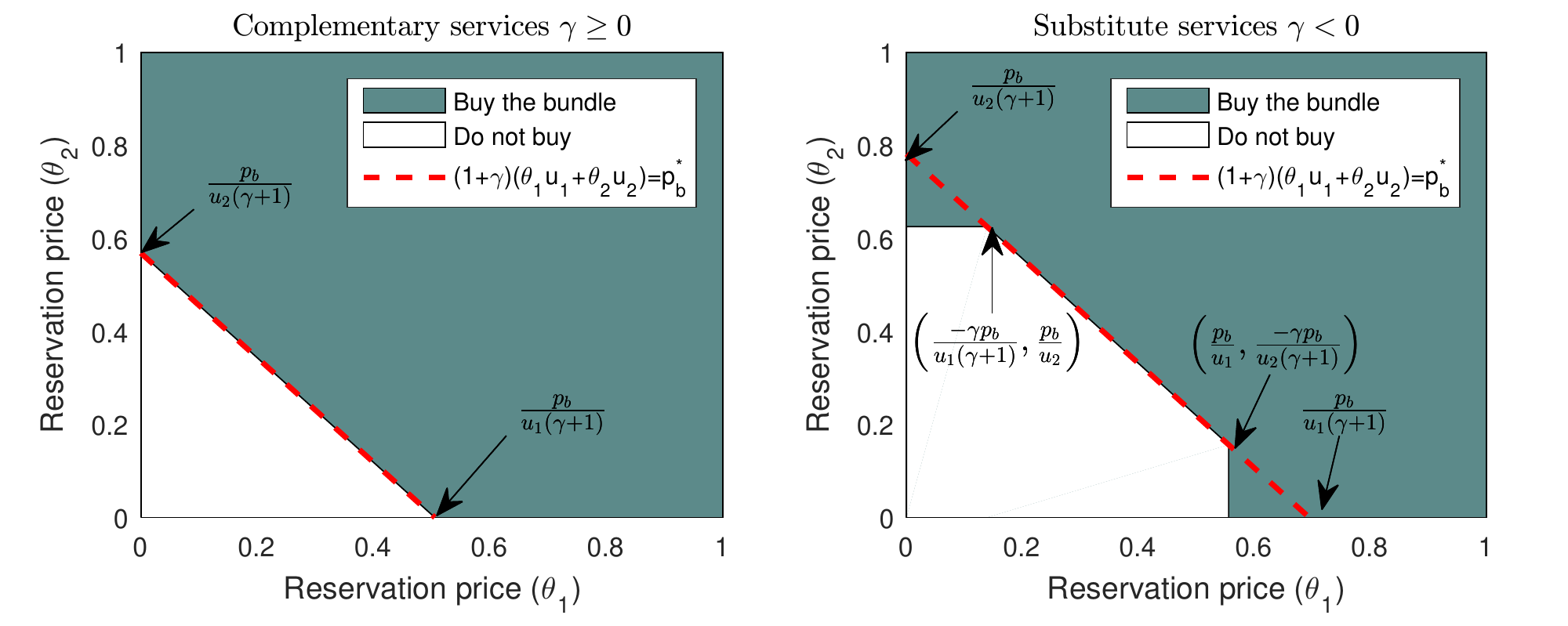}
\par\end{centering}
\caption{Customer demand on the service bundles of complements and substitutes.\label{fig:bundling_cases}}
\end{figure*}

\subsection{Market Model and Degree of Contingency}

We consider the marketing strategy of virtually bundling two services denoted as service~$S_{1}$ and service~$S_{2}$ into a bundle denoted as service bundle $S_{b}$ as shown in Figure~\ref{fig:bundle_system}. We identify the degree of contingency between the two services as $\gamma$. $\gamma$ indicates the customer interest in obtaining the two services $S_{1}$ and $S_{2}$ as a service bundle $S_{b}$. We incorporate the definition of $\gamma$ from microeconomics as follows~\cite{venkatesh2003optimal}:
\begin{equation}
\gamma=\frac{\theta_{b}-(\theta_{1}+\theta_{2})}{\theta_{1}+\theta_{2}},\label{eq:degree_contingency}
\end{equation}
where $\theta_{b}$, $\theta_{1}$, and $\theta_{2}$ are the reservation prices of the service bundle~$S_{b}$, the standalone service~$S_{1}$, and the standalone service~$S_{2}$, respectively. The service bundle $S_{b}$ can be classified into two types as follows:
\begin{itemize}
\item $\theta_{b}\geq(\theta_{1}+\theta_{2})$ and hence $\gamma\geq0$ : When $\theta_{b}$ is greater than or equal to the summation of $\theta_{1}$ and $\theta_{2}$, $S_{1}$ and $S_{2}$ are complementary services. For example, the customers are willing to buy both services~$S_{1}$ and $S_{2}$ as each service has a unique functionality.
\item $\theta_{b}<(\theta_{1}+\theta_{2})$ and hence $\gamma<0$: When $\theta_{b}$ is less than the summation of $\theta_{1}$ and $\theta_{2}$, $S_{1}$ and $S_{2}$ are substitute services. For example, the customers are not willing to buy both services~$S_{1}$ and $S_{2}$ as they are similar and comparable in functionality.
\end{itemize}
The customer demand on buying $S_{b}$ can be found as in Figure~\ref{fig:bundling_cases}. $u_{k}$ is the service quality of the $k$-th service in $S_{b}$. The shaded regions represent the probability of buying $S_{b}$. $p_{b}$ is the subscription fee of $S_{b}$. Each point in the figure represents the reservation price pair $(\theta_{1},\theta_{2})$ of the two services in the bundle. For both complementary and substitute services, a customer will buy $S_{b}$ if $(\theta_{1},\theta_{2})$ lies above the decision line $(1+\gamma)(\theta_{1}u_{1}+\theta_{2}u_{2})=p_{b}$. Moreover, the customers of substitute services will also buy $S_{b}$ when $(\theta_{1},\theta_{2})$ lies at the right side of the line $\frac{p_{b}}{u_{1}}$ and above the line $\frac{p_{b}}{u_{2}}$. We observe that the customers are more willing to buy $S_{b}$ containing complementary services than substitute services.

\subsection{Complementary People-Centric Services ($\gamma\geq0$)}

$S_{1}$ and $S_{2}$ are complements when the reservation price of the service bundle $S_{b}$ is greater than the total reservation price of the standalone services $\theta_{b}\geq(\theta_{1}+\theta_{2})$. The gross profit of $S_{b}$ containing complementary services can be defined as follows:
\begin{multline}
G_{c}(r_{1},r_{2},p_{b})=\underbrace{Mp_{b}\mathbb{P}\left((1+\gamma)(\theta_{1}u_{1}+\theta_{2}u_{2})>p_{b}\right)}_{\text{Total subscription revenue}}\\
-\underbrace{N\left(c_{1}\left(1-r_{1}\right)+c_{2}\left(1-r_{2}\right)\right)}_{\text{Total data cost of }S_{1}\text{and }S_{2}},\label{eq:complementary_fun_1}
\end{multline}
where $M$ is the number of potential customers willing to buy the service bundle $S_{b}$. $p_{b}$ is the subscription fee of $S_{b}$. $r_{1}$ and $r_{2}$ are the privacy levels of $S_{1}$ and $S_{2}$, respectively. $N$ is the number of crowdsensing participants. $c_{1}$ and $c_{2}$ are the reservation wages of the crowdsensing participants in $S_{1}$ and $S_{2}$, respectively. $u_{1}=\alpha_{1}-\alpha_{2}\exp\left(\alpha_{3}r_{1}\right)$ is the service quality of $S_{1}$ and $u_{2}=\beta_{1}-\beta_{2}\exp\left(\beta_{3}r_{2}\right)$ is the service quality of $S_{2}$ as formulated in (\ref{eq:accuracy-privacy_fun}). $\mathbb{P}\left((1+\gamma)(\theta_{1}u_{1}+\theta_{2}u_{2})>p_{b}\right)$ is the probability of buying $S_{b}$ as defined by the shaded area of complementary services in Figure~\ref{fig:bundling_cases}. Assuming that $\theta_{1}$ and $\theta_{2}$ follow a uniform distribution, (\ref{eq:complementary_fun_1}) can be re-written as follows:
\begin{multline}
G_{c}(r_{1},r_{2},p_{b})=Mp_{b}\left(1-\frac{0.5p_{b}^{2}}{\left(1+\gamma\right)^{2}u_{1}u_{2}}\right)\\
-Nc_{1}\left(1-r_{1}\right)-Nc_{2}\left(1-r_{2}\right).\label{eq:complementary_fun}
\end{multline}

The profit maximization problem by selling two complements in $S_{b}$ is then expressed as follows:
\begin{equation}
\begin{aligned}\underset{r_{1},r_{2},p_{b}}{\text{maximize }} & G_{c}(r_{1},r_{2},p_{b})\\
\text{subject to } & C3:p_{b}\geq0,\\
 & C4:r_{1}\geq0,\\
 & C5:r_{2}\geq0,
\end{aligned}
\label{eq:complements_optimization}
\end{equation}
where $C3$, $C4$, and $C5$ are the optimization constraints required to ensure nonnegative solutions of $p_{b}$, $r_{1}$, and $r_{2}$, respectively. The Lagrangian dual function of (\ref{eq:complements_optimization}) is derived as:
\begin{equation}
\mathcal{L}_{c}\left(r_{1},r_{2},p_{b},\lambda,\lambda_{4},\lambda_{5}\right)=G_{c}(r_{1},r_{2},p_{b})+\lambda_{3}p_{b}+\lambda_{4}r_{1}+\lambda_{5}r_{2},\label{eq:complements_optimization_lagrange}
\end{equation}
where $\lambda_{3}$, $\lambda_{4}$, and $\lambda_{5}$ are the Lagrange multipliers.
\begin{prop}
\label{prop:complements_solutions}The closed-form solutions $p_{b}^{*}$, $r_{1}^{*}$, and $r_{2}^{*}$ of (\ref{eq:complements_optimization}) are given as follows:
\begin{equation}
p_{b}^{*}=\frac{0.5A_{1}}{M\alpha_{3}\beta_{3}},\label{eq:complements_p_b_optimal}
\end{equation}
\begin{multline}
r_{1}^{*}=\frac{1}{\alpha_{3}}\log\bigg(\frac{13.5N^{2}c_{1}c_{2}}{M^{2}\alpha_{2}\alpha_{3}\beta_{1}\beta_{3}\left(\gamma^{2}+2\gamma+1\right)}\\
+\frac{2.25Nc_{1}A_{1}}{M^{2}\alpha_{2}\alpha_{3}^{2}\beta_{1}\beta_{3}\left(\gamma^{2}+2\gamma+1\right)}\bigg),\label{eq:complements_r1_optimal}
\end{multline}
\begin{multline}
r_{2}^{*}=\frac{1}{\beta_{3}}\log\bigg(\frac{13.5N^{2}c_{1}c_{2}}{M^{2}\alpha_{1}\alpha_{3}\beta_{2}\beta_{3}\left(\gamma^{2}+2\gamma+1\right)}\\
+\frac{2.25Nc_{2}A_{1}}{M^{2}\alpha_{1}\alpha_{3}\beta_{2}\beta_{3}^{2}\left(\gamma^{2}+2\gamma+1\right)}\bigg),\label{eq:complements_r2_optimal}
\end{multline}

\end{prop}
where
\begin{multline}
A_{1}=\big[\frac{8}{2}\alpha_{1}\beta_{1}M^{2}\alpha_{3}^{2}\gamma^{2}\beta_{3}^{2}+\frac{16}{3}\alpha_{1}\beta_{1}M^{2}\alpha_{3}^{2}\gamma\beta_{3}^{2}\\
+\frac{8}{2}\alpha_{1}\beta_{1}M^{2}\alpha_{3}^{2}\beta_{3}^{2}+9N^{2}\alpha_{3}^{2}c_{2}^{2}-18N^{2}\alpha_{3}c_{1}c_{2}\beta_{3}\\
+9N^{2}c_{1}^{2}\beta_{3}^{2}\big]^{0.5}-3N\alpha_{3}c_{2}-3N\beta_{3}c_{1}.
\end{multline}

\begin{IEEEproof}
This result can be proved by taking the first derivatives of (\ref{eq:complements_optimization_lagrange}) with respect to $r_{1}$, $r_{2}$, and $p_{b}$. From the three resulting equations and with inactive constraints ($\lambda_{1}=\lambda_{2}=\lambda_{3}=0$), the closed form solutions can be obtained.
\end{IEEEproof}
\begin{prop}
\label{prop:complements_concavity}The profit function $G_{c}(r_{1},r_{2},p_{b})$ defined in (\ref{eq:complementary_fun}) for complementary people-centric services is concave. The closed-form solutions $p_{b}^{*}$, $r_{1}^{*}$, and $r_{2}^{*}$ given in (\ref{eq:complements_p_b_optimal}), (\ref{eq:complements_r1_optimal}), and (\ref{eq:complements_r2_optimal}), respectively, are globally optimal.
\end{prop}
\begin{IEEEproof}
The Hessian matrix $\mathbf{H}_{G}$ of $G_{c}(r_{1},r_{2},p_{b})$ is found as in (\ref{eq:complements_hessian_matrix}), shown at the top of the next page.
\begin{figure*}[t]
\begin{equation}
\mathbf{H}_{G}=\begin{bmatrix}-\frac{0.5M\alpha_{2}\alpha_{3}^{2}p_{b}^{3}exp\left(\alpha_{3}r_{1}\right)u_{1}}{\left(\gamma+1\right)^{2}u_{1}^{3}u_{2}} & -\frac{0.5M\alpha_{2}\alpha_{3}p_{b}^{3}\beta_{2}\beta_{3}exp\left(\alpha_{3}r_{1}\right)exp\left(r_{2}\beta_{3}\right)}{\left(\gamma+1\right)^{2}u_{1}^{2}u_{2}^{2}} & -\frac{1.5M\alpha_{2}\alpha_{3}p_{b}^{2}exp\left(\alpha_{3}r_{1}\right)}{\left(\gamma+1\right)^{2}u_{1}^{2}u_{2}}\\
-\frac{0.5M\alpha_{2}\alpha_{3}p_{b}^{3}\beta_{2}\beta_{3}exp\left(\alpha_{3}r_{1}\right)exp\left(r_{2}\beta_{3}\right)}{\left(\gamma+1\right)^{2}u_{1}^{2}u_{2}^{2}} & -\frac{0.5Mp_{b}^{3}\beta_{2}\beta_{3}^{2}exp\left(r_{2}\beta_{3}\right)u_{2}}{\left(\gamma+1\right)^{2}u_{1}u_{2}^{3}} & -\frac{1.5Mp_{b}^{2}\beta_{2}\beta_{3}exp\left(r_{2}\beta_{3}\right)}{\left(\gamma+1\right)^{2}u_{1}u_{2}^{2}}\\
-\frac{1.5M\alpha_{2}\alpha_{3}p_{b}^{2}exp\left(\alpha_{3}r_{1}\right)}{\left(\gamma+1\right)^{2}u_{1}^{2}u_{2}} & -\frac{1.5Mp_{b}^{2}\beta_{2}\beta_{3}exp\left(r_{2}\beta_{3}\right)}{\left(\gamma+1\right)^{2}u_{1}u_{2}^{2}} & -\frac{3Mp_{b}}{\left(\gamma+1\right)^{2}u_{1}u_{2}}
\end{bmatrix}.
\label{eq:complements_hessian_matrix}
\end{equation}
\end{figure*}
The leading principal minors can then be obtained as in (\ref{eq:complements_pri_minors_1})-(\ref{eq:complements_pri_minors_3}), shown at the top of the next page,
\begin{figure*}[t]
\begin{eqnarray}
D_{1} & = & -\frac{0.5M\alpha_{2}\alpha_{3}^{2}p_{b}^{3}exp\left(\alpha_{3}r_{1}\right)\left(\alpha_{1}+\alpha_{2}exp\left(\alpha_{3}r_{1}\right)\right)}{\left(\gamma+1\right)^{2}\left(\beta_{1}-\beta_{2}exp\left(\beta_{3}r_{2}\right)\right)\left(\alpha_{1}-\alpha_{2}exp\left(\alpha_{3}r_{1}\right)\right)^{3}},\label{eq:complements_pri_minors_1}\\
D_{2} & = & \frac{0.25M^{2}\alpha_{2}\alpha_{3}^{2}p_{b}^{6}\beta_{2}\beta_{3}^{2}exp\left(\alpha_{3}r_{1}+r_{2}\beta_{3}\right)\left(\alpha_{1}\beta_{1}+\alpha_{2}\beta_{1}exp\left(\alpha_{3}r_{1}\right)+\alpha_{1}\beta_{2}exp\left(r_{2}\beta_{3}\right)\right)}{\left(\gamma+1\right)^{4}\left(\beta_{1}-\beta_{2}exp\left(r_{2}\beta_{3}\right)\right)^{4}\left(\alpha_{1}-\alpha_{2}exp\left(\alpha_{3}r_{1}\right)\right)^{4}},\label{eq:complements_pri_minors_2}\\
D_{3} & = & \frac{0.375A_{2}}{\left(\gamma+1\right)^{6}\left(\beta_{1}-\beta_{2}exp\left(\beta_{3}r_{2}\right)\right)^{5}\left(\alpha_{1}-\alpha_{2}exp\left(\alpha_{3}r_{1}\right)\right)^{5}}\label{eq:complements_pri_minors_3}.
\end{eqnarray}
\end{figure*}
where 
\begin{multline}
A_{2}=M^{3}\alpha_{1}\alpha_{2}\alpha_{3}^{2}p_{b}^{7}\beta_{2}^{2}\beta_{3}^{2}exp\left(\alpha_{3}r_{1}\right)exp\left(2\beta_{3}r_{2}\right)\\
+M^{3}\alpha_{2}^{2}\alpha_{3}^{2}p_{b}^{7}\beta_{1}\beta_{2}\beta_{3}^{2}exp\left(\beta_{3}r_{2}\right)exp\left(2\alpha_{3}r_{1}\right)\\
-2M^{3}\alpha_{1}\alpha_{2}\alpha_{3}^{2}p_{b}^{7}\beta_{1}\beta_{2}\beta_{3}^{2}exp\left(\alpha_{3}r_{1}\right)exp\left(\beta_{3}r_{2}\right).
\end{multline}
$G_{c}(r_{1},r_{2},p_{b})$ is concave as $D_{1}\leq0$, $D_{2}\geq0$, and $D_{3}\leq0$ when $A_{2}\leq0$. Accordingly, the closed-form solutions are globally optimal.
\end{IEEEproof}

We next present the optimal solutions to the profit maximization problem with fixed privacy levels for $S_{1}$ and $S_{2}$.

\subsubsection{Fixed Privacy Level}

When the privacy levels are enforced by an external legislation entity, we have the following proposition.
\begin{prop}
When the privacy levels $r_{1}$ and $r_{2}$ are fixed by an external entity, the optimal subscription fee of the service bundle is expressed as follows:
\begin{equation}
p_{b}^{*}=\frac{0.82\left(\gamma+1\right)\left(\alpha_{1}-\alpha_{2}exp\left(\alpha_{3}r_{1}\right)\right)\left(\beta_{1}-\beta_{2}exp\left(\beta_{3}r_{2}\right)\right)}{\left(\left(\alpha_{1}-\alpha_{2}exp\left(\alpha_{3}r_{1}\right)\right)\left(\beta_{1}-\beta_{2}exp\left(\beta_{3}r_{2}\right)\right)\right)^{0.5}}.\label{eq:p_b_optimal_fixed_privacy}
\end{equation}
(\ref{eq:p_b_optimal_fixed_privacy}) is globally optimal.
\end{prop}
\begin{IEEEproof}
The second derivatives of $G_{c}(r_{1},r_{2},p_{b})$ defined in (\ref{eq:complementary_fun}) with respect to $p_{b}$ is derived as follows:
\begin{multline}
\frac{\partial^{2}G_{c}\left(\cdot\right)}{\partial p_{b}^{2}}=\\
\frac{-3Mp_{b}}{\left(\alpha_{1}-\alpha_{2}exp\left(\alpha_{3}r_{1}\right)\right)\left(\beta_{1}-\beta_{2}exp\left(r_{2}\beta_{3}\right)\right)\left(\gamma+1\right)^{2}}\leq0,
\end{multline}
which is nonpositive. Thus, the profit maximization problem with a fixed privacy level is concave and the solution in (\ref{eq:p_b_optimal_fixed_privacy}) is globally optimal.
\end{IEEEproof}

\subsection{Substitute People-Centric Service ($\gamma<0$)}

As they have comparable functionality, substitute services are only required to obtain better overall service quality, e.g., predictive performance. Services~$S_{1}$ and $S_{2}$ are called substitutes when the reservation price of the service bundle $S_{b}$ is less than the total reservation price of the separate sales $\theta_{b}<(\theta_{1}+\theta_{2})$. Bundling substitute services yields in the following gross profit:
\begin{multline}
G_{c}(r_{1},r_{2},p_{b})=\underbrace{Mp_{b}\mathbb{P}\left(\begin{array}{c}
\left[(1+\gamma)(\theta_{1}u_{1}+\theta_{2}u_{2})>p_{b}\right]\\
\cup\left[\theta_{1}\geq\frac{p_{b}}{u_{1}}\right]\cup\left[\theta_{2}\geq\frac{p_{b}}{u_{2}}\right]
\end{array}\right)}_{\text{Total subscription revenue}}\\
-\underbrace{Nc_{1}\left(1-r_{1}\right)-Nc_{2}\left(1-r_{2}\right)}_{\text{Total data cost of }S_{1}\text{and }S_{2}}.\label{eq:substitute_fun_1}
\end{multline}
$\mathbb{P}\left(\left[(1+\gamma)(\theta_{1}u_{1}+\theta_{2}u_{2})>p_{b}\right]\cup\left[\theta_{1}\geq\frac{p_{b}}{u_{1}}\right]\cup\left[\theta_{2}\geq\frac{p_{b}}{u_{2}}\right]\right)$ is the probability for a customer to buy $S_{b}$ which can be defined by the shaded area of substitute services in Figure~\ref{fig:bundling_cases}. Then, (\ref{eq:substitute_fun_1}) can be re-written as follows: 
\begin{multline}
G_{s}(r_{1},r_{2},p_{b})=Mp_{b}\left(1-\frac{0.5p_{b}^{2}+\gamma^{2}p_{b}^{2}}{\left(1+\gamma\right)^{2}u_{1}u_{2}}\right)\\
-Nc_{1}\left(1-r_{1}\right)-Nc_{2}\left(1-r_{2}\right).\label{eq:substitute_fun}
\end{multline}
The profit maximization problem of selling two substitute services in $S_{b}$ is expressed as follows:
\begin{equation}
\begin{aligned}\underset{r_{1},r_{2},p_{b}}{\text{maximize }} & G_{s}(r_{1},r_{2},p_{b})\\
\text{subject to } & C6:p_{b}\geq0,\\
 & C7:r_{1}\geq0,\\
 & C8:r_{2}\geq0.
\end{aligned}
\label{eq:substitute_optimization}
\end{equation}
The objective is maximizing the gross profit of $S_{b}$ under the constraints $C6$, $C7$, and $C8$ for nonnegative solutions in $p_{b}^{*}$, $r_{1}^{*}$, and $r_{2}^{*}$, respectively.
\begin{prop}
The profit function $G_{s}(r_{1},r_{2},p_{b})$ defined in (\ref{eq:substitute_fun}) for substitute people-centric services is concave. The closed-form solutions $p_{b}^{*}$, $r_{1}^{*}$, and $r_{2}^{*}$ are given in~(\ref{eq:substitute_p_b_optimal}), (\ref{eq:substitute_r1_optimal}), and (\ref{eq:substitute_r1_optimal}), respectively,
\begin{figure*}[t]
\begin{eqnarray}
p_{b}^{*} & = & -\frac{0.5A_{3}}{M\alpha_{3}\beta_{3}},\label{eq:substitute_p_b_optimal}\\
r_{1}^{*} & = & \frac{1}{\alpha_{3}}\log\left(\frac{13.5\left(c_{1}c_{2}N^{2}\gamma^{2}+c_{1}c_{2}N^{2}\right)}{M^{2}\alpha_{2}\alpha_{3}\beta_{1}\beta_{3}\left(\gamma^{2}+2\gamma+1\right)}-\frac{2.25\left(Nc_{1}\gamma^{2}+Nc_{1}\right)A_{3}}{M^{2}\alpha_{2}\alpha_{3}^{2}\beta_{1}\beta_{3}\left(\gamma^{2}+2\gamma+1\right)}\right),\label{eq:substitute_r1_optimal}\\
r_{2}^{*} & = & \frac{1}{\beta_{3}}\log\left(\frac{13.5Nc_{2}\left(Nc_{1}\gamma^{2}+Nc_{1}\right)}{M^{2}\alpha_{1}\alpha_{3}\beta_{2}\beta_{3}\left(\gamma^{2}+2\gamma+1\right)}-\frac{2.25Nc_{2}\left(\gamma^{2}+1\right)\left(A_{3}\right)}{M^{2}\alpha_{1}\alpha_{3}\beta_{2}\beta_{3}^{2}\left(\gamma^{2}+2\gamma+1\right)}\right).\label{eq:substitute_r2_optimal}
\end{eqnarray}
\end{figure*}

\end{prop}
where 
\begin{multline}
A_{3}=3N\alpha_{3}c_{2}+3Nc_{1}\beta_{3}-3N\alpha_{3}c_{2}-3Nc_{1}\beta_{3}\\
+\frac{1}{9\left(\gamma^{2}+1\right)^{2}}\big[8\alpha_{1}\beta_{1}M^{2}\alpha_{3}^{2}\gamma^{2}\beta_{3}^{2}+16\alpha_{1}\beta_{1}M^{2}\alpha_{3}^{2}\gamma\beta_{3}^{2}\\
+8\alpha_{1}\beta_{1}M^{2}\alpha_{3}^{2}\beta_{3}^{2}+27N^{2}\alpha_{3}^{2}c_{2}^{2}\gamma^{2}+27N^{2}\alpha_{3}^{2}c_{2}^{2}\\
-54N^{2}\alpha_{3}c_{1}c_{2}\gamma^{2}\beta_{3}-54N^{2}\alpha_{3}c_{1}c_{2}\beta_{3}+27N^{2}c_{1}^{2}\gamma^{2}\beta_{3}^{2}\\
+27N^{2}c_{1}^{2}\beta_{3}^{2}\big]^{0.5}.
\end{multline}
These closed-form solutions are globally optimal.
\begin{IEEEproof}
The proof is similar to the ones of Propositions~\ref{prop:complements_solutions} and \ref{prop:complements_concavity} and will be omitted due to the space limit.
\end{IEEEproof}

\subsection{Profit Sharing\label{sub:profit_sharing}}

A service bundle can be formed by two service providers forming a bundling coalition $\mathcal{K}$. We next present a profit sharing model to divide the bundling profit among the cooperative providers.  

\subsubsection{Core Solution}

Let $\varphi_{k}$ indicate the profit share of the service provider $S_{k}$, where $k\in\mathcal{K}$. The core solution $\mathcal{C}$ is defined as follows~\cite{myerson2013game}:
\begin{equation}
\mathcal{C}=\bigg\{\mathbf{\varphi}\mid\underbrace{\sum_{k\in\mathcal{K}}\varphi_{k}=G_{\mathcal{K}}^{*}}_{\text{group rationality}}\text{ and }\underbrace{\sum_{k\in\mathcal{S}}\varphi_{k}\geq F_{\mathcal{S}}^{*},\mathcal{S}\subseteq\mathcal{K}}_{\text{individual rationality}}\bigg\}\label{eq:core_sol}
\end{equation}
where $G_{\mathcal{K}}^{*}$ is the bundling profit and $F_{\mathcal{S}}^{*}$ is the profit resulting from selling the services separately.

The core solution $\mathcal{C}$ can contain an infinite number of possible share allocations, can be empty, or lead to unfair share allocations when considering the contributions of services in $S_{b}$. We next present the Shapley value concept which provides a fair and single solution to the profit sharing problem of $S_{b}$.

\subsubsection{The Shapley Value Solution}

For each service $S_{k}$, where $k\in\mathcal{K},$ forming the bundle $S_{b}$, the Shapley value solution $\eta=\left(\eta_{1},\eta_{2}\right)$ ensures fairness and assigns a payoff $\eta_{k}$ defined as~\cite{myerson2013game}:
\begin{equation}
\eta_{k}=\sum_{\mathcal{S}\subseteq\mathcal{K}\setminus\left\{ k\right\} }\underbrace{\frac{\left|\mathcal{S}\right|!\left(\left|\mathcal{K}\right|-\left|\mathcal{S}\right|-1\right)!}{\left|\mathcal{K}\right|!}}_{\text{probability of random ordering}}\underbrace{\left(G_{\mathcal{K}}^{*}-F_{\mathcal{S}}^{*}\right)}_{\text{marginal contribution}}.\label{eq:shapley_sol}
\end{equation}
The first term defines the random order of joining the bundle. The second term defines the marginal contribution of each service on increasing the bundling profit.

\section{Experimental Results\label{sec:experimental_results}}

In this section, we first present three people-centric services which are trained using real-world datasets. We also analyze the quality of the services when deep learning~\cite{Goodfellow2016deep} and random forests~\cite{breiman2001random} are utilized as data analytics algorithms. We then introduce numerical results of selling the services separately. Finally, we evaluate the bundling models for selling complementary and substitute services, respectively.

\subsection{People-Centric Services and Bundles}

Using real-world datasets, we design the following people-centric services:
\begin{itemize}
\item \textbf{\emph{Service~}}\emph{$S_{1}$ (sentiment analysis using deep learning)}: Using the Sentiment140 dataset~\cite{go2009twitter}, we develop a service to predict people's sentiment from social networking tweets. The sentiment can be either positive or negative. People post tweets which typically include personal information, and privacy awareness is reasonably required. We use $629,145$ tweet samples for model training and $419,431$ tweet samples for model testing and quality calculation. We assume that the reservation wage of each crowdsensing participant is $0.2$\footnote{We use monetary units for all payment, cost, and profit analysis in the experimental results. Actual currency, such as the United States dollar, can be applied without affecting the optimization models or results.}.
\item \textbf{\emph{Service~}}\emph{$S_{2}$ (sentiment analysis using random forests)}: This service is similar to service\emph{~$S_{1}$} above, except in using a random forest classifier instead of deep learning.
\item \textbf{\emph{Service~}}\emph{$S_{3}$ (activity tracking using random forests)}: This service enables the tracking of human activities using the accelerometer sensors of mobile devices. We use the Actitracker dataset~\cite{kwapisz2011activity} containing a time series of $1,098,207$~data points. We divide the accelerometer time series into overlapping window frames resulting in $23,072$~training and $5,768$~testing samples. The predicted activities are walking, jogging, walking upstairs, walking downstairs, sitting, and standing. We set the reservation wage of each crowdsensing participant as $0.1$.
\end{itemize}
These people-centric services can be sold separately or interrelated in service bundles. We consider the following bundling scenarios:
\begin{itemize}
\item \textbf{\emph{Bundle~}}\emph{$S_{b1}$ (}$S_{1}$ and $S_{3}$\emph{)}: Section~\ref{sub:exp_complements} considers the economic strategy of virtually packaging services~$S_{1}$ and $S_{3}$ into one service bundle. Services~$S_{1}$ and $S_{3}$ are complementary as both services are typically required to provide in-depth understanding of mobile users. We assume that the degree of contingency is $\gamma=0.1$ which indicates the high customer willingness in acquiring both services at once.
\item \textbf{\emph{Bundle~}}\emph{$S_{b2}$ (}$S_{1}$ and $S_{2}$\emph{)}: In Section~\ref{sub:exp_substitute}, we analyze services~$S_{1}$ and $S_{2}$ as substitutes because they have comparable functionality, i.e., both services~$S_{1}$ and $S_{2}$ are used for sentiment analysis, but they differ in the data analytics algorithm. To reflect the low customer willingness of buying comparable services, the degree of contingency is set as $\gamma=-0.1$.
\end{itemize}
We set the number of crowdsensing participants to $N=100$ and the number of customers to $M=1000$.

\begin{figure*}
\begin{centering}
\includegraphics[width=1\linewidth,trim=0cm 0cm 0cm 0cm]{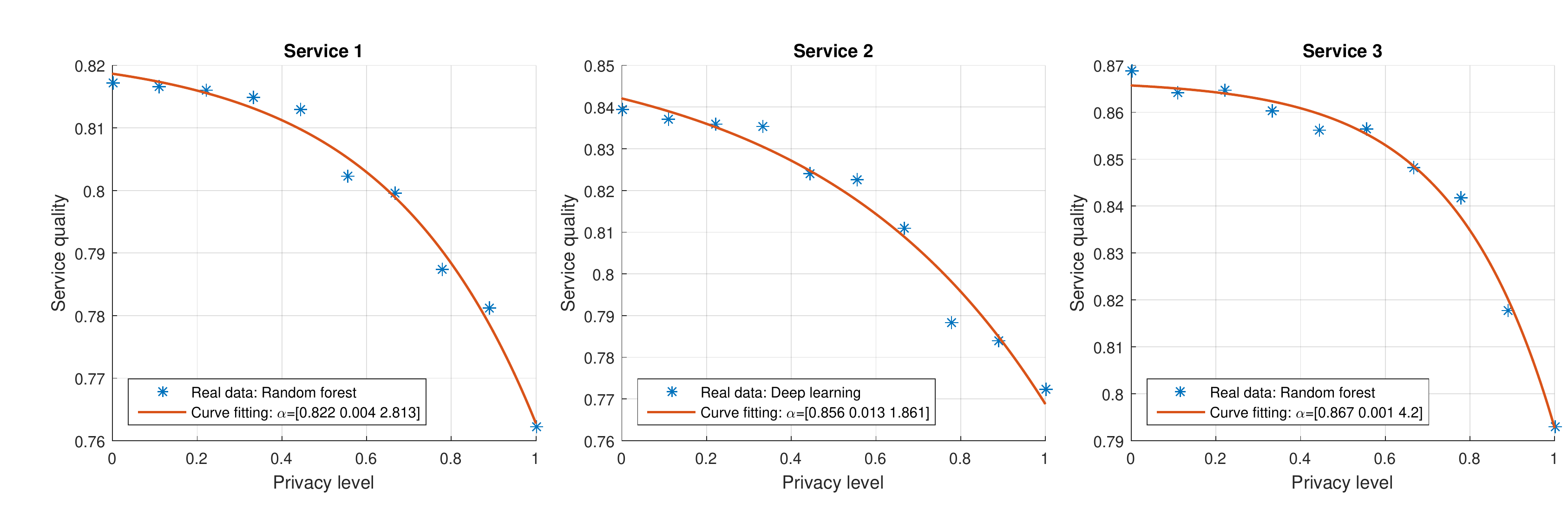}
\par\end{centering}
\caption{The prediction quality of the services $S_{1}$, $S_{2}$, and $S_{3}$ (from left to right) under varied privacy levels.\label{fig:utility_privacy}}
\end{figure*}

\subsection{The Quality-Privacy Tradeoff\label{sub:accuracy_privacy_tradeoff}}

Figure~\ref{fig:utility_privacy} shows the quality-privacy models of $S_{1}$, $S_{2}$, and $S_{3}$, respectively. We observe three major results. Firstly, the service quality decreases as the privacy level increases. This is clear as increasing the privacy level results in higher data distortion. Secondly, it can be also noted that the real data fits the quality function defined in~(\ref{eq:accuracy-privacy_fun}). Thirdly, the service quality of $S_{1}$ and $S_{2}$ are different even though they use the same dataset. This is due to the different data analytics algorithms used in $S_{1}$ and $S_{2}$.

\subsection{Standalone Sales}

We use service~$S_{1}$ to evaluate the profit maximization model for selling people-centric services as a standalone product. From Figure~\ref{fig:utility_privacy}, the quality-privacy fitting parameters of $S_{1}$ are $\alpha_{1}=0.822$, $\alpha_{2}=0.004$, and $\alpha_{3}=2.813$.

\begin{figure}
\begin{centering}
\includegraphics[width=1\columnwidth]{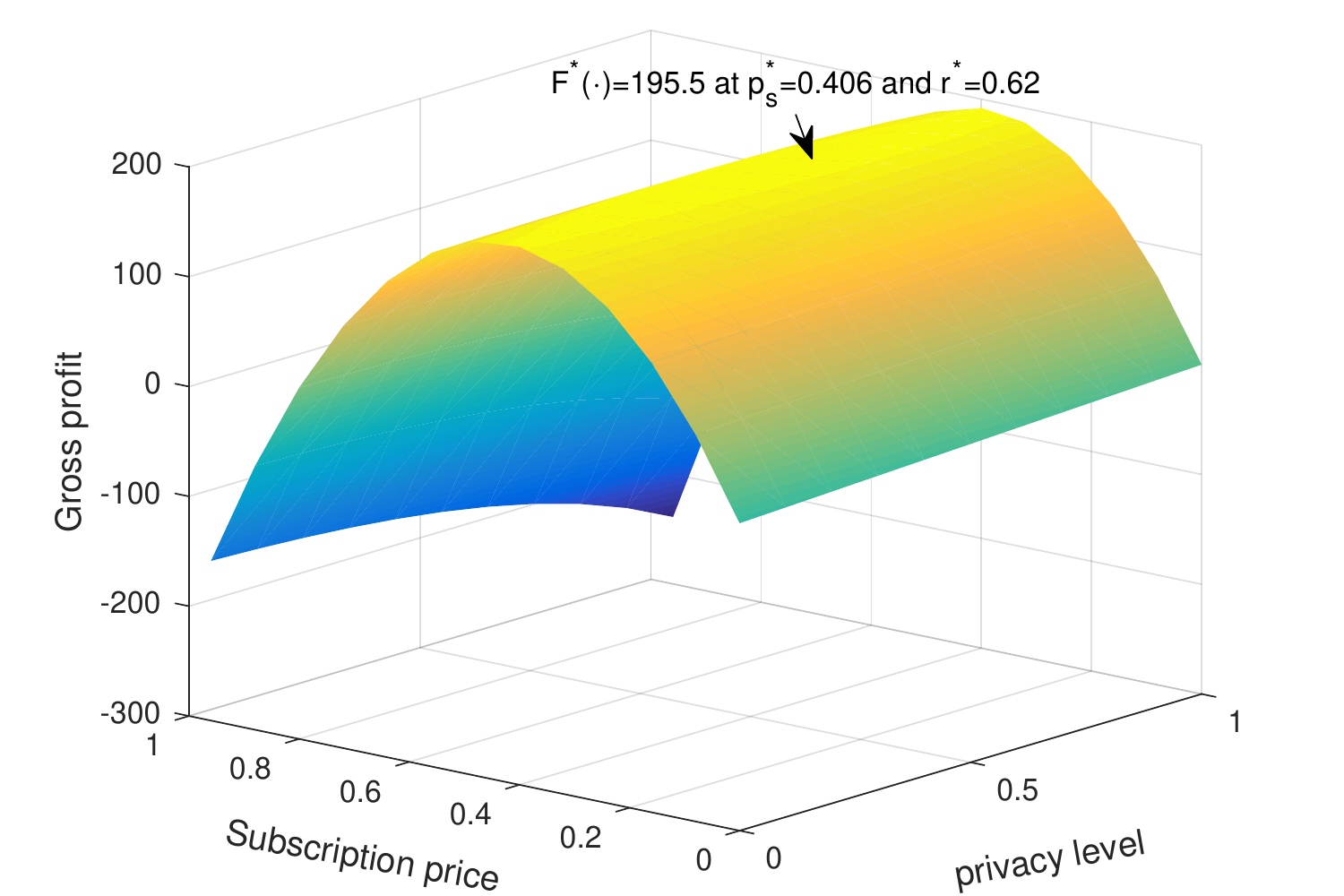}
\par\end{centering}
\caption{Gross profit of $S_{1}$ under varied privacy level $r$ and subscription fee $p_{s}$.\label{fig:seperate_selling}}
\end{figure}

\subsubsection{Gross Profit Optimization}

Figure~\ref{fig:seperate_selling} shows the gross profit $F^{*}(r,p_{s})$ of $S_{1}$ defined in (\ref{eq:profit_function_seperate}) under varied privacy level $r$ and subscription fee $p_{s}$. When the subscription fee is high, the profit decreases as fewer customers will be willing to pay the subscription fee. When the subscription fee is low, more customers will buy $S_{1}$. However, the gross profit falls down due to the low subscription fee. Likewise, a high privacy level results in a low service quality and fewer customers will be accordingly interested in the service of poor quality. A low privacy level results in a high service quality, but gross profit will decrease due to the high spending in buying the true data from the crowdsensing participants. The optimal settings of the subscription fee $p_{s}^{*}=0.406$ and privacy level $r^{*}=0.62$ can be found using the closed-form solutions in (\ref{eq:p_s_optimal}) and (\ref{eq:r_optimal}), respectively. Then using (\ref{eq:profit_function_seperate}), the maximum profit is calculated as $F(r^{*},p_{s}^{*})=195.5$.

\begin{figure}
\begin{centering}
\includegraphics[width=1\columnwidth]{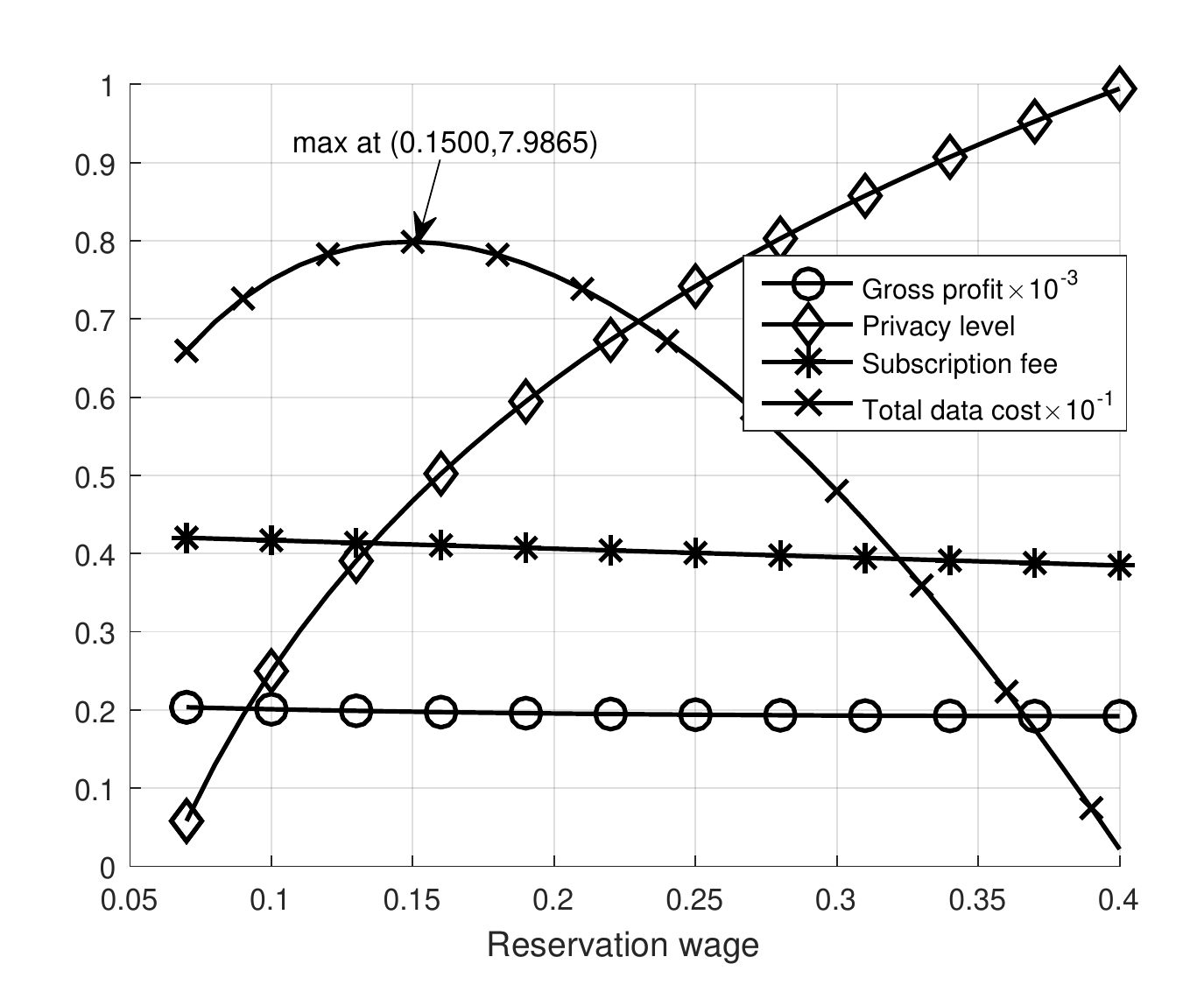}
\par\end{centering}
\caption{Impacts of the reservation wage $c$ on the gross profit $F^{*}(\cdot)$, privacy level~$r^{*}$, subscription fee~$p_{s}^{*}$, and total data cost.\label{fig:seperate_reservation_wage}}
\end{figure}

\begin{figure}
\begin{centering}
\includegraphics[width=1\columnwidth]{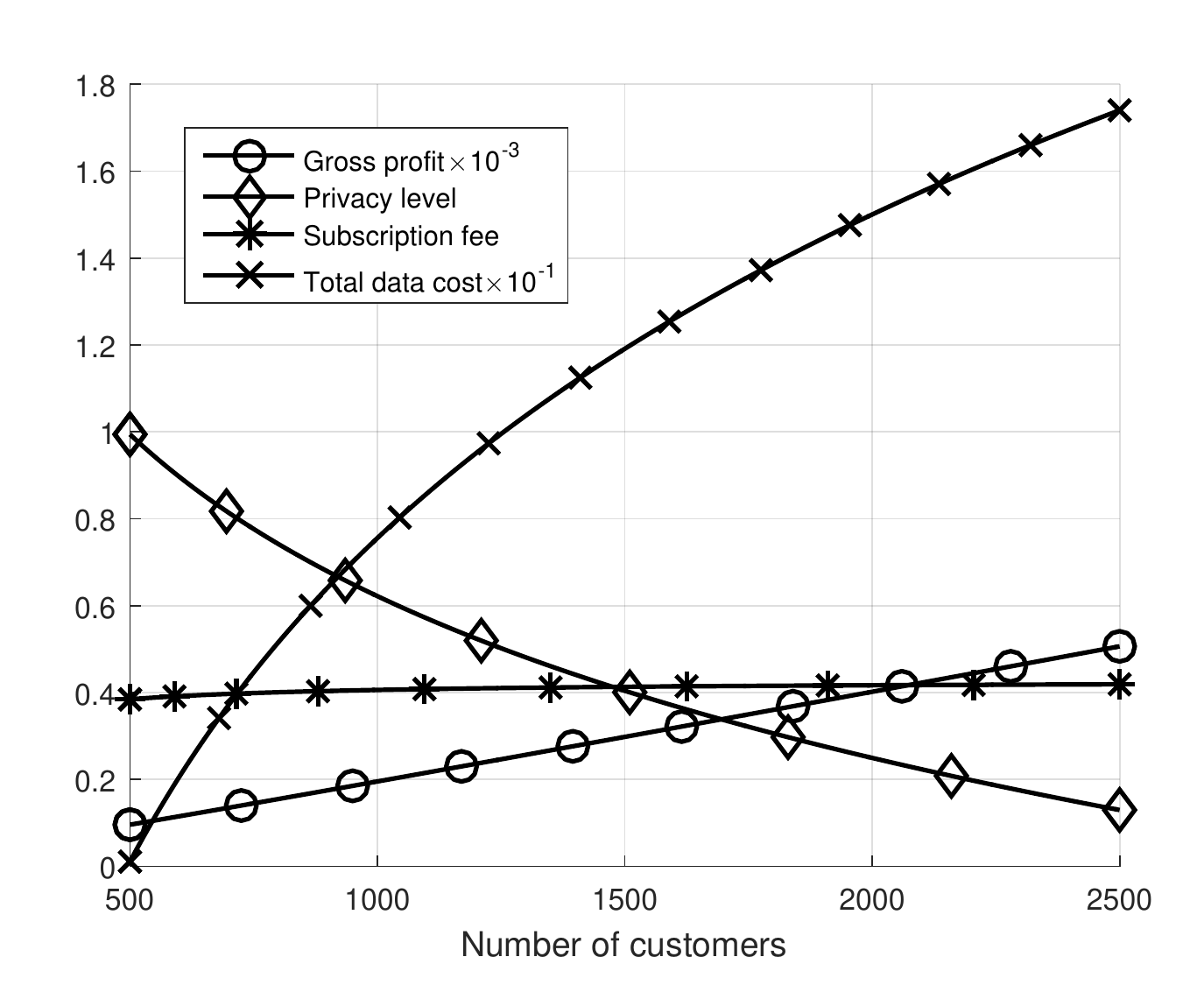}
\par\end{centering}
\caption{Impacts of the customer base size $M$ on the gross profit $F^{*}(\cdot)$, privacy level~$r^{*}$, subscription fee~$p_{s}^{*}$, and total data cost.\label{fig:seperate_customers}}
\end{figure}

\subsubsection{The Impact of Reservation Wage}

In Figure~\ref{fig:seperate_reservation_wage}, we consider the impact of varying the reservation wage of the crowdsensing participants on the gross profit $F^{*}(\cdot)$, privacy level~$r^{*}$, subscription fee~$p_{s}^{*}$, and total data cost. Firstly, there is an inverse correlation between the reservation wage and the gross profit. Specifically, when the reservation wage is increased, the total data cost will increase up to $c=0.15$ and the profit will accordingly decrease. Increasing the reservation wage beyond $c=0.15$ yields a fall of the total data price as a rational service provider will intensively increase the privacy level as defined in (\ref{eq:r_optimal}). Secondly, we note that the reservation wage and subscription fee are also inversely proportional. In particular, the service provider reduces the subscription fee to attract more customers due to the degradation in the service quality.

\subsubsection{The Impact of Customer Base}

Figure~\ref{fig:seperate_customers} shows the gross profit $F^{*}(\cdot)$, privacy level~$r^{*}$, subscription fee~$p_{s}^{*}$, and total data cost under varied number of customers. When the number of customers increases, the gross profit and subscription fee increase as the benefit of the increased demand. Moreover, the service provider decreases the privacy level to collect more true data which increases the service quality and total data cost.

\begin{figure}
\begin{centering}
\includegraphics[width=1\columnwidth]{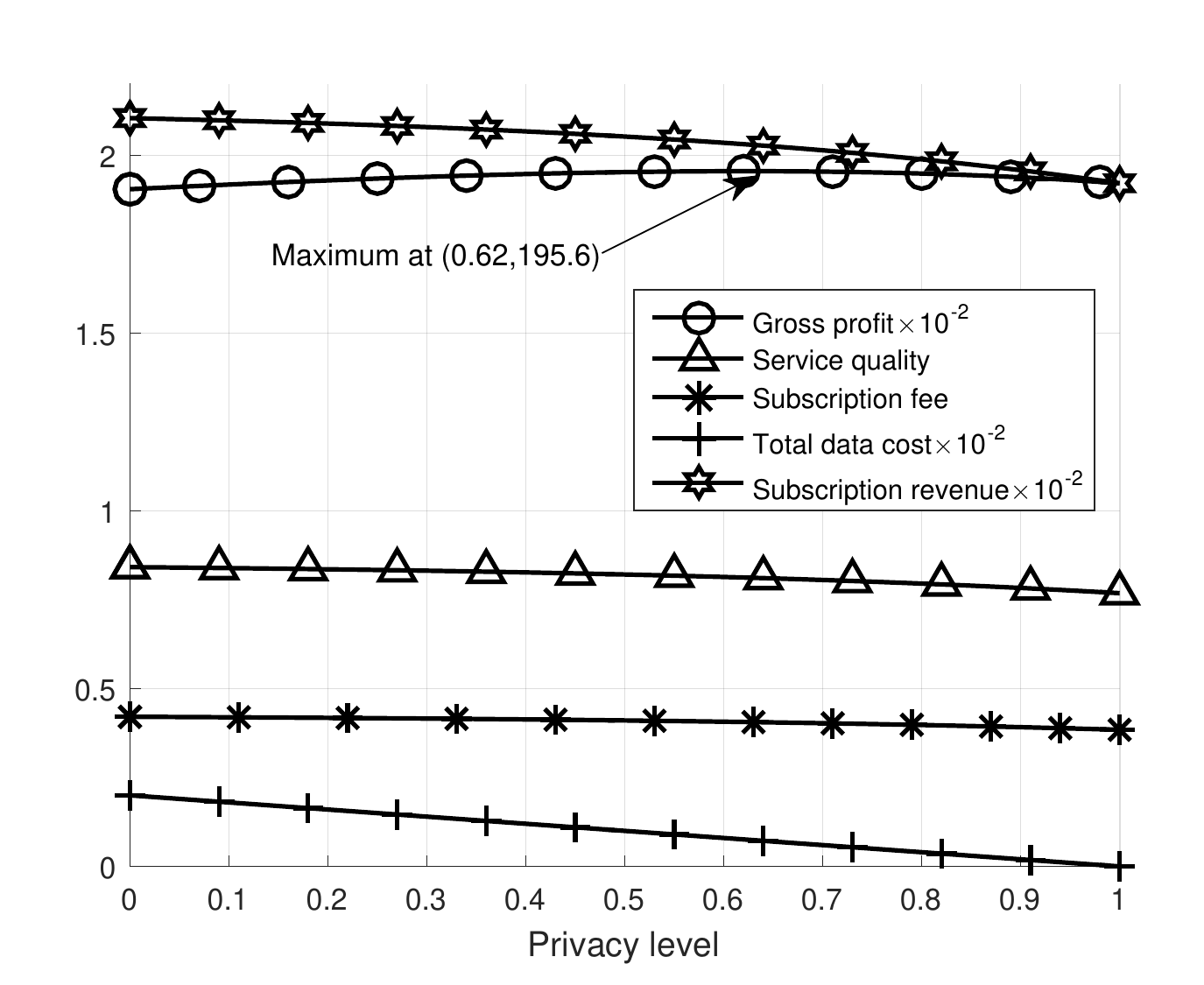}
\par\end{centering}
\caption{Gross profit $F^{*}(\cdot)$, subscription revenue, subscription fee~$p_{s}^{*}$, and total data cost under varied privacy level~$r$.\label{fig:seperate_fixed_privacy}}
\end{figure}

\subsubsection{Fixed Privacy Level}

In some scenarios, the service provider does not control the privacy level as discussed in Section~\ref{sub:seperate_fixed_privacy}, e.g., due to legislation rules. Instead, the service provider only specifies the subscription fee as in (\ref{eq:p_s_optimal_fixed_privacy}) to gain the maximum gross profit. In Figure~\ref{fig:seperate_fixed_privacy}, we analyze the gross profit $F^{*}(\cdot)$, subscription revenue, subscription fee~$p_{s}^{*}$, and total data cost of $S_{1}$ at varied privacy level~$r$. We observe two important results. Firstly, the subscription revenue, subscription fee, and total data cost are inversely correlated with the privacy level. This is expected as increasing the privacy level negatively affects the service quality and fewer customers will be interested in buying $S_{1}$. Besides, the total data cost will decrease when the privacy level is high. Secondly, we note that the gross profit increases up to $r=0.62$, then it decreases due to the extreme loss of customers at the high privacy levels $r>0.62$.

\subsection{Complementary People-Centric Services\label{sub:exp_complements}}

We consider bundling $S_{1}$ and $S_{3}$ as complementary services into $S_{b1}$. From Figure~\ref{fig:utility_privacy}, the fitting parameters of $S_{1}$ are $\alpha_{1}=0.822$, $\alpha_{2}=0.004$, and $\alpha_{3}=2.813$. For $S_{3}$, the fitting parameters are $\beta_{1}=0.867$, $\beta_{2}=0.001$, and $\beta_{3}=4.2$. We first analyze the bundling profit and the impacts of the different parameters on $S_{b1}$. We then present the payoff allocations among $S_{1}$ and $S_{3}$ based on the importance of each service on the sales of $S_{b1}$. 

\begin{figure*}
\begin{centering}
\subfloat[]{\begin{centering}
\includegraphics[width=1\columnwidth]{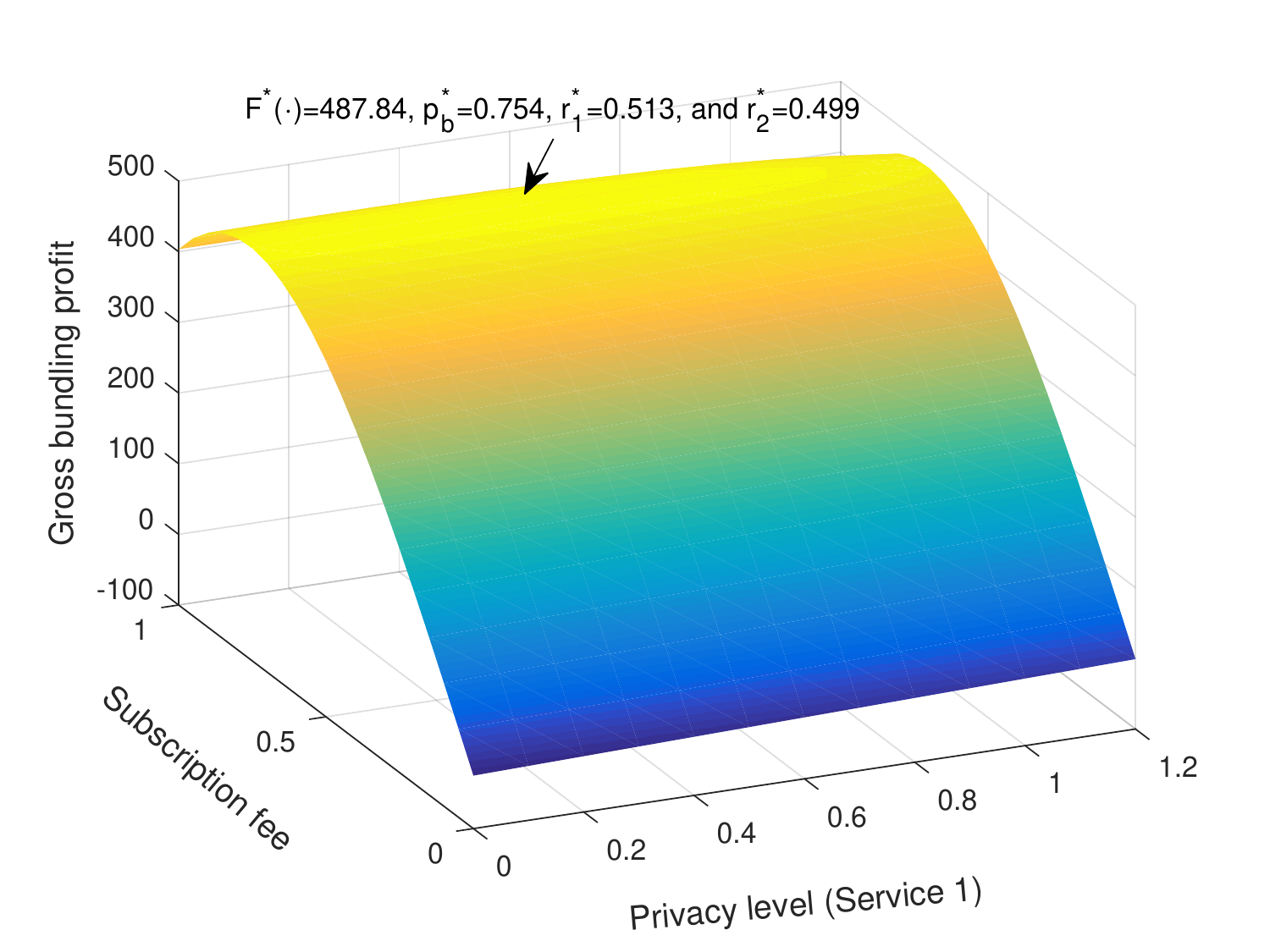}
\par\end{centering}

}\subfloat[]{\begin{centering}
\includegraphics[width=1\columnwidth]{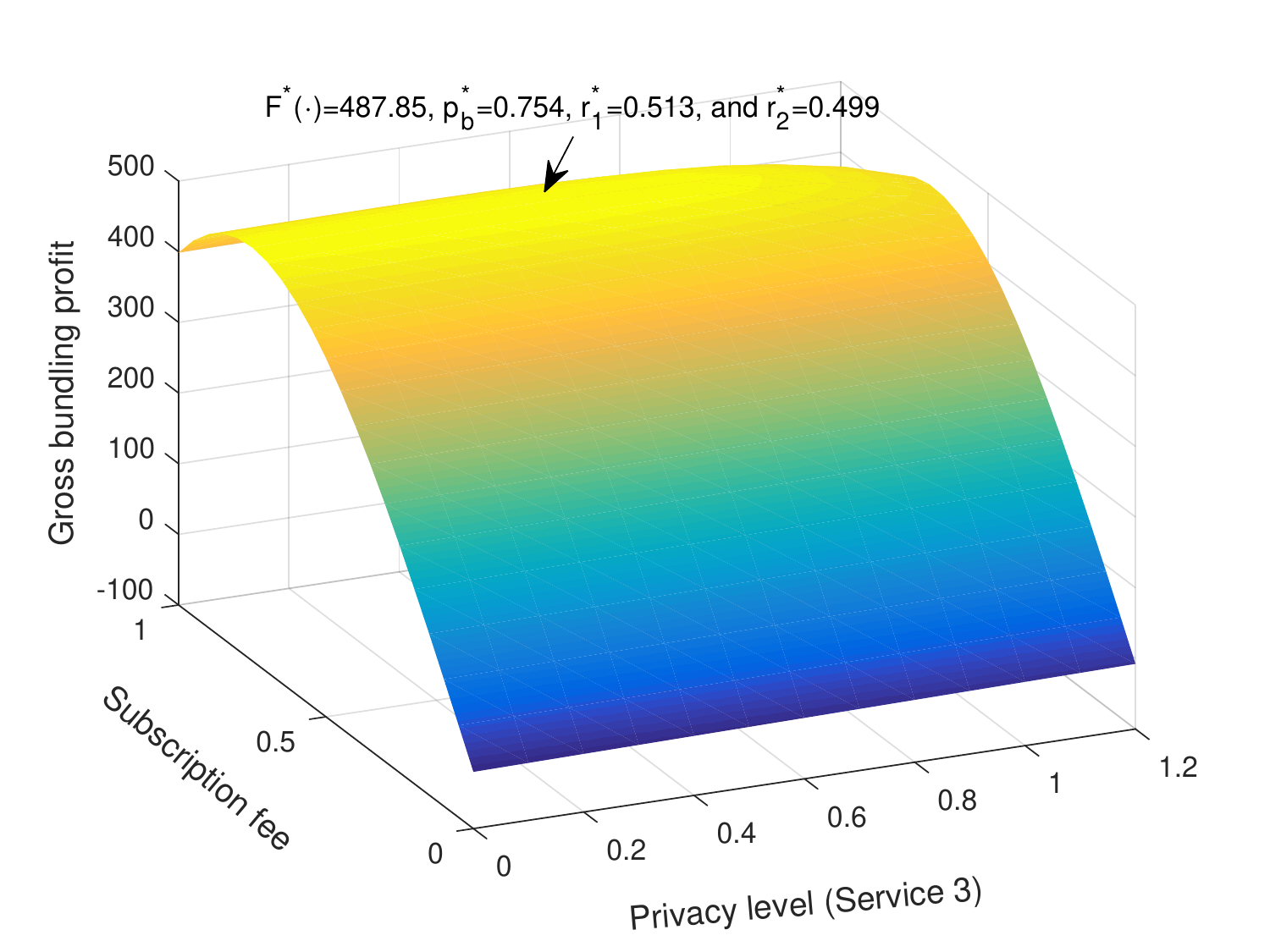}
\par\end{centering}

}
\par\end{centering}

\caption{Gross profit $G_{c}(r_{1},r_{2},p_{b})$ by bundling $S_{1}$ and $S_{3}$ into the service bundle $S_{b1}$ under varied privacy levels $r_{1}$ and $r_{2}$ and subscription fee $p_{b}$.\label{fig:complements_selling}}
\end{figure*}

\subsubsection{Gross Profit Optimization}

The gross bundling profit $G_{c}(r_{1},r_{2},p_{b})$ defined in (\ref{eq:complementary_fun}) is presented in Figure~\ref{fig:complements_selling}. When the subscription fee $p_{b}$ and the privacy levels $r_{1}$ and $r_{2}$ are either high or low, the gross profit goes down. Specifically, fewer customers will buy overpriced or poor quality service bundles. Likewise, $S_{b1}$ makes a low profit when the subscription fee and privacy level are low due to the low revenue and high total data cost, respectively. The optimal settings $p_{b}^{*}=0.754$, $r_{1}^{*}=0.513$, and $r_{2}^{*}=0.499$ can be obtained using the closed-form solutions given in (\ref{eq:complements_p_b_optimal}), (\ref{eq:complements_r1_optimal}), and (\ref{eq:complements_r2_optimal}), respectively. The optimal gross profit of $S_{b1}$ is $G_{c}(r_{1}^{*},r_{2}^{*},p_{b}^{*})=487.84$ which is greater than that of selling services~$S_{1}$ and $S_{3}$ as standalone products with $F_{1}(r^{*},p_{s}^{*})=195.5$ and $F_{3}(r^{*},p_{s}^{*})=206.02$, respectively. Thus, the rational service providers will decide to build $S_{b1}$ and stop selling $S_{1}$ and $S_{3}$ as standalone services.

\begin{figure}
\begin{centering}
\includegraphics[width=0.9\columnwidth]{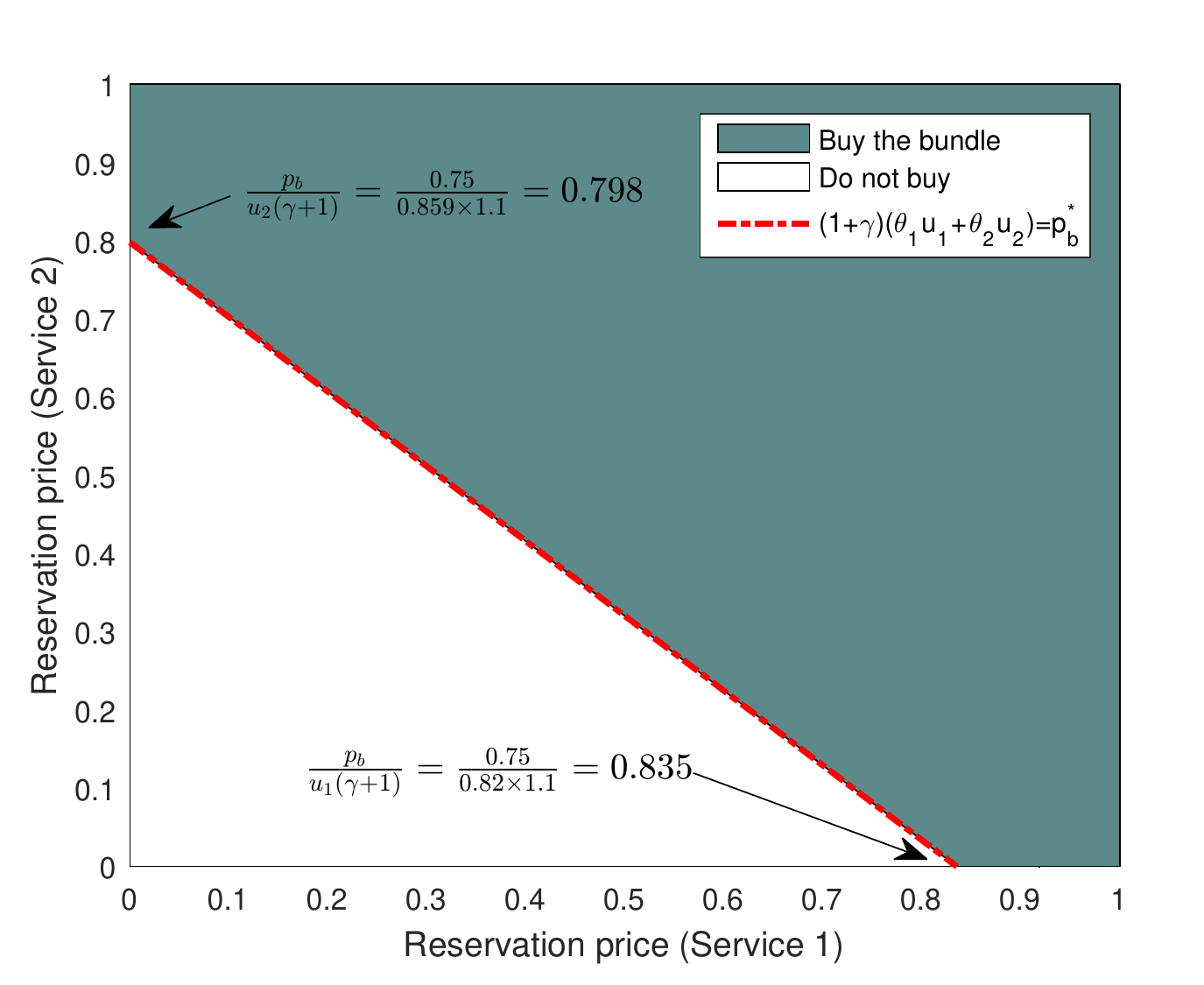}
\par\end{centering}
\caption{Demands on $S_{b1}$ containing both $S_{1}$ and $S_{3}$.\label{fig:complements_decision}}
\end{figure}

\subsubsection{Demand Boundaries}

Figure~\ref{fig:complements_decision} shows the demand boundary of $S_{b1}$ in the reservation price spaces. The customers buy $S_{b1}$ when the customer valuations lie above the decision line $(1+\gamma)(\theta_{1}u_{1}+\theta_{2}u_{2})=p_{b}^{*}$, where $u_{1}=0.82$, $u_{2}=0.859$, $\gamma=0.1$, and $p_{b}^{*}=0.754$. The customers do not buy $S_{b1}$ when their valuations are below the decision line.

\begin{figure}
\begin{centering}
\includegraphics[width=1\columnwidth]{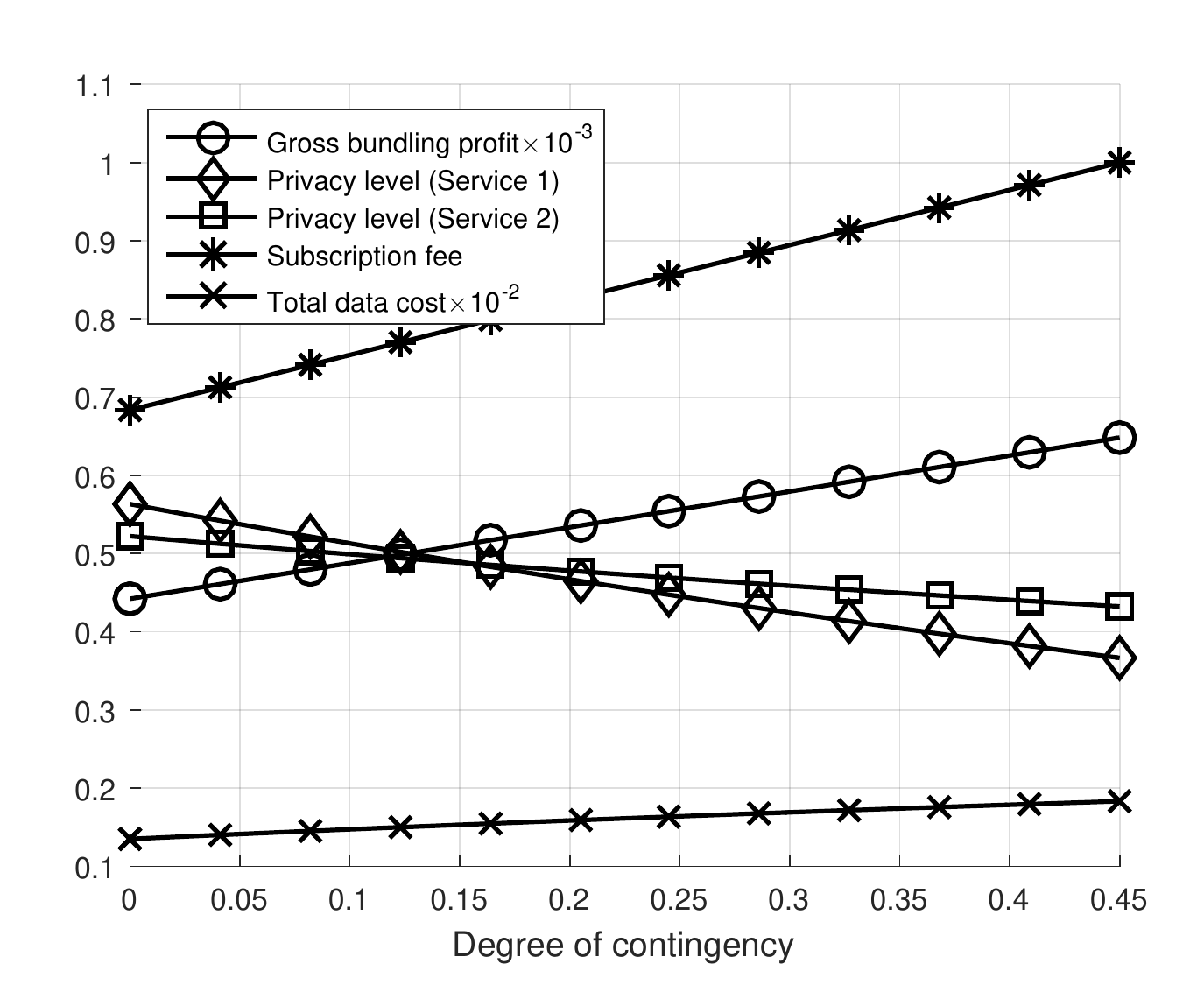}
\par\end{centering}
\caption{Impacts of the degree of contingency~$\gamma$ on the gross profit $G_{c}^{*}(\cdot)$, privacy levels~$r_{1}^{*}$ and $r_{2}^{*}$, subscription fee~$p_{b}^{*}$, and total data cost.\label{fig:complements_contingency}}
\end{figure}

\begin{figure}
\begin{centering}
\includegraphics[width=1\columnwidth]{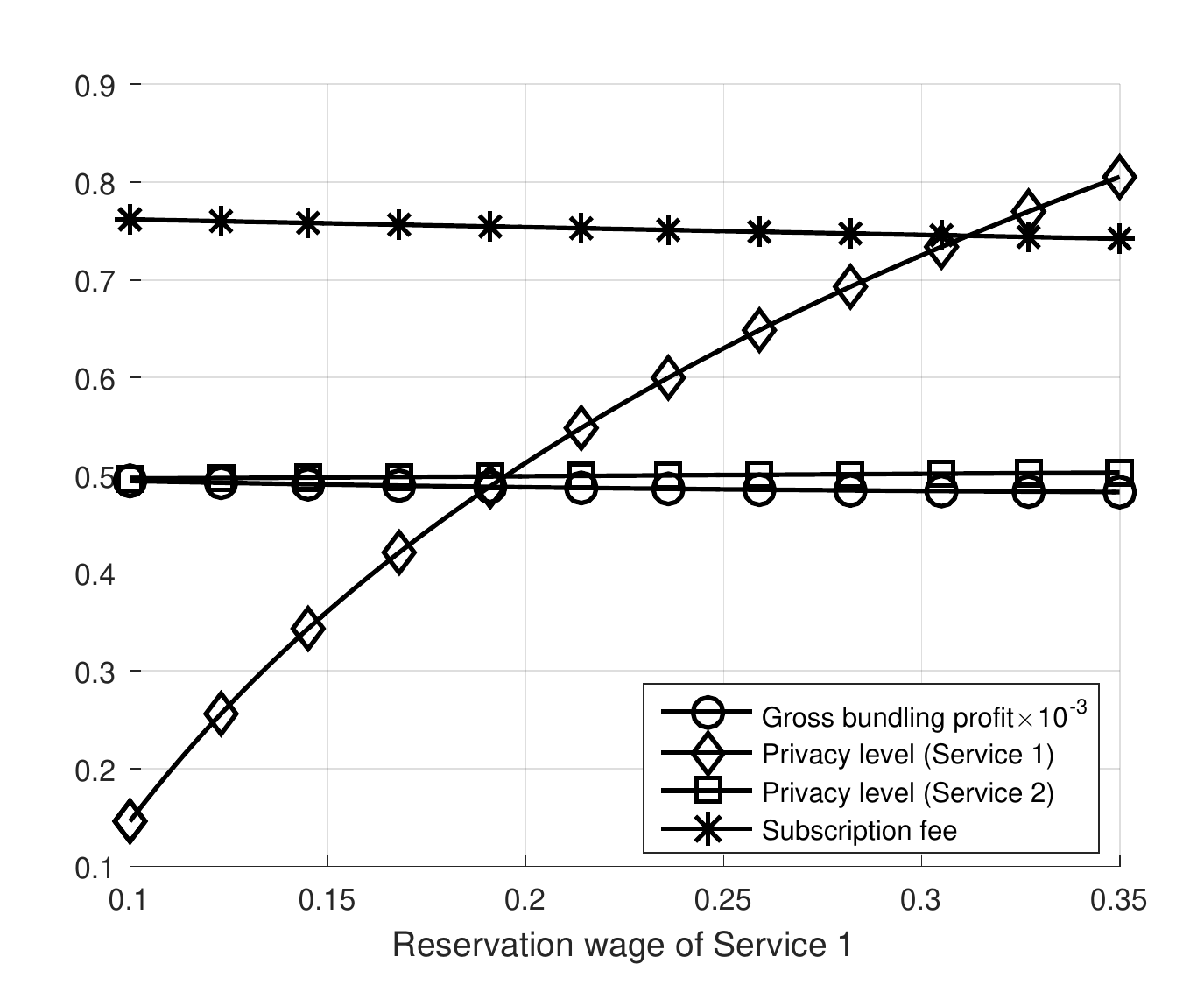}
\par\end{centering}
\caption{Impacts of the reservation wage $c_{1}$ on the gross profit $G_{c}^{*}(\cdot)$, privacy levels~$r_{1}^{*}$ and $r_{2}^{*}$, and subscription fee~$p_{b}^{*}$.\label{fig:complements_reservation wage}}
\end{figure}

\subsubsection{The Impact of Contingency Degree}

In Figure~\ref{fig:complements_contingency}, we analyze the gross profit $G_{c}^{*}(\cdot)$, privacy levels~$r_{1}^{*}$ and $r_{2}^{*}$, subscription fee~$p_{b}^{*}$, and total data cost under varied degree of contingency $\gamma$. The total data cost of service bundle~$S_{b1}$ includes the data costs of $S_{1}$ and $S_{3}$ as expressed in (\ref{eq:complementary_fun}). Firstly, we note that the gross bundling profit is proportional to the degree of contingency. This is clear as the high degree of contingency indicates strong interrelation between $S_{1}$ and $S_{3}$. Thus, the customers are more interested in buying both services together. Secondly, the subscription fee of $S_{b1}$ is increased to meet any increase in the degree of contingency. The resulting increase in the gross profit motivates the service provider to enhance the overall service quality by decreasing the privacy levels $r_{1}^{*}$ and $r_{2}^{*}$.

\subsubsection{The Impact of Reservation Wage}

We consider the impact of varying the reservation wage of service~$S_{1}$ on the optimal pricing and profits of service bundle~$S_{b1}$. We observe two important results from Figure~\ref{fig:complements_reservation wage}. Firstly, the bundling profit $G_{c}^{*}(\cdot)$ goes down when the reservation wage $c_{1}$ increases. This is due to the increased data cost of $S_{1}$ as defined in (\ref{eq:complementary_fun}). Secondly, in order to minimize the total data cost, the privacy level $r_{1}^{*}$ of $S_{1}$ is increased. The privacy level $r_{2}^{*}$ of $S_{2}$ is also slightly increased but at a lower rate than $r_{1}^{*}$. These results can also be deduced from the closed-form solutions of $r_{1}^{*}$ and $r_{2}^{*}$ in (\ref{eq:complements_r1_optimal}) and (\ref{eq:complements_r2_optimal}), respectively.

\begin{figure}
\begin{centering}
\includegraphics[width=0.9\columnwidth]{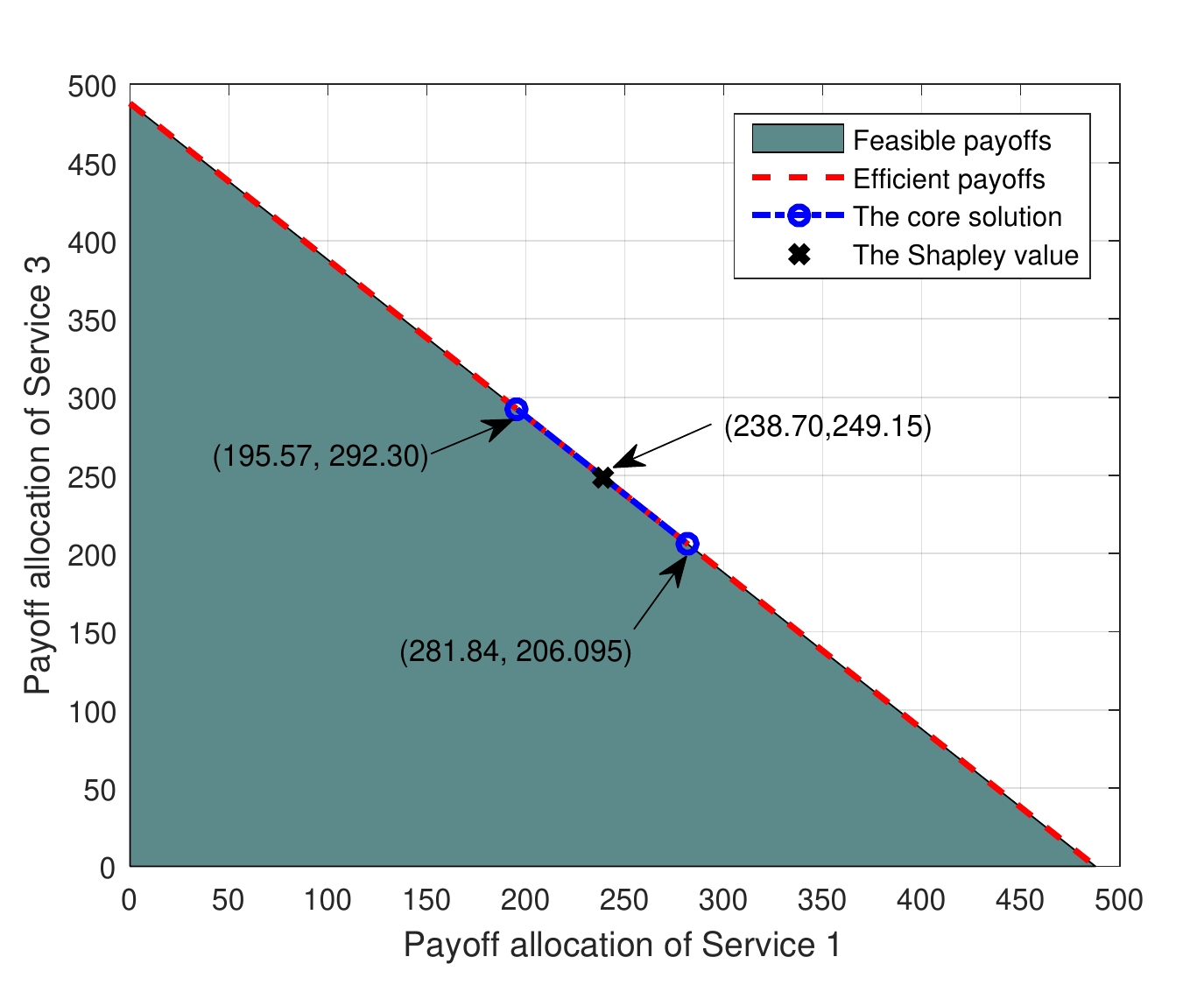}
\par\end{centering}
\caption{Payoff allocation of the bundling profit from $S_{b1}$ among $S_{1}$ and $S_{3}$.\label{fig:complements_profit_sharing}}
\end{figure}

\begin{figure}
\begin{centering}
\includegraphics[width=1\columnwidth]{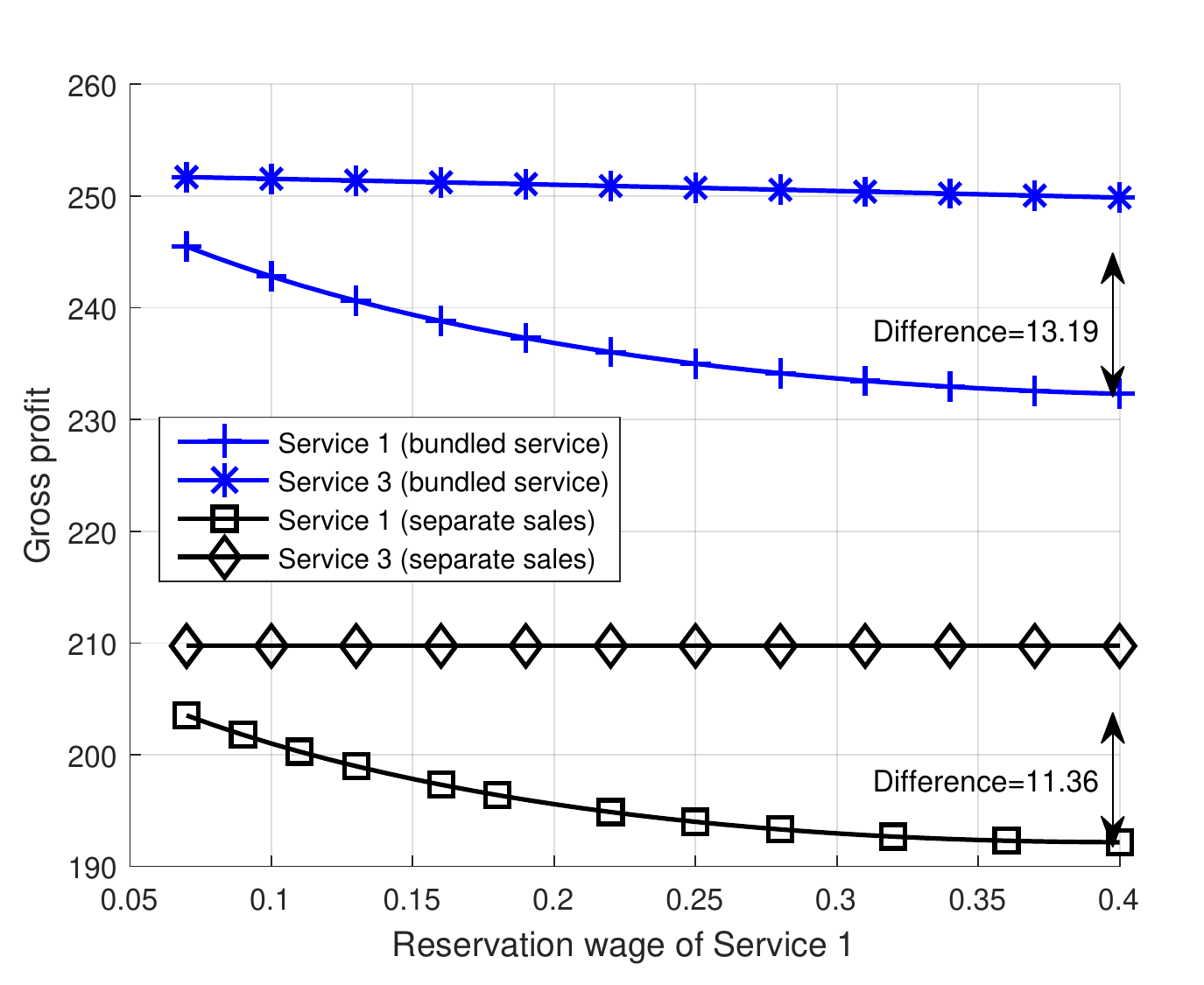}
\par\end{centering}
\caption{Payoff allocation under varied reservation wage $c_{1}$ of $S_{1}$. \label{fig:complements_profit_sharing_wage}}
\end{figure}

\subsubsection{Profit Sharing}

The bundling profit can be divided between services~$S_{1}$ and $S_{3}$ as shown in Figure~\ref{fig:complements_profit_sharing}. The feasible payoffs guarantee that the summation of payoffs does not exceed the bundling profit $\eta_{1}+\eta_{3}\leq G_{c}(r_{1}^{*},r_{2}^{*},p_{b}^{*})=487.84$. The efficient payoffs assign the allocations such that the total payoff is equal to the bundling profit $\eta_{1}+\eta_{3}=G_{c}(r_{1}^{*},r_{2}^{*},p_{b}^{*})=487.84$. The core solution defined in (\ref{eq:core_sol}) ensures that the payoff allocations of either $S_{1}$ or $S_{3}$ cannot be improved by leaving the bundle and selling services separately. Finally, the Shapley value solution defined in (\ref{eq:shapley_sol}) assigns fair payoff allocations based on the importance of each service forming $S_{b1}$.

We next study the impacts of varying the reservation wage $c_{1}$ of $S_{1}$ on the profit shares from $S_{b1}$. Figure~\ref{fig:complements_profit_sharing_wage} shows the profit resulting from offering $S_{1}$ and $S_{3}$ separately and jointly as $S_{b1}$. The profit allocations in $S_{b1}$ is found using the Shapley value solution defined in (\ref{eq:shapley_sol}). Two important observations can be made. Firstly, the gross profit falls as $c_{1}$ is increased. This has negative effects on the profit of $S_{3}$ in both the bundling and separate sales. The maximum-to-minimum profit difference of $S_{3}$ in the bundling and separate sales are $13.19$ and $11.36$, respectively. Secondly, we observe that higher profit allocations can be obtained from $S_{b1}$ for $S_{1}$ and $S_{3}$ compared to the separate sales. Thus, the providers of services~$S_{1}$ and $S_{3}$ have a monetary incentive in making the service bundle~$S_{b1}$ regardless of the data cost. This result shows that the fairness of the Shapley value solution is crucial for stable service bundling in people-centric services.

\subsection{Substitute People-Centric Services\label{sub:exp_substitute}}

As substitute services, we next consider combining services~$S_{1}$ and $S_{2}$ in into the service bundle~$S_{b2}$. As shown in Figure~\ref{fig:utility_privacy}, the fitting parameters of $S_{1}$ are $\alpha_{1}=0.822$, $\alpha_{2}=0.004$, and $\alpha_{3}=2.813$. For $S_{2}$, the fitting parameters are $\beta_{1}=0.856$, $\beta_{2}=0.013$, and $\beta_{3}=1.861$.

\begin{figure}
\begin{centering}
\includegraphics[width=0.9\columnwidth]{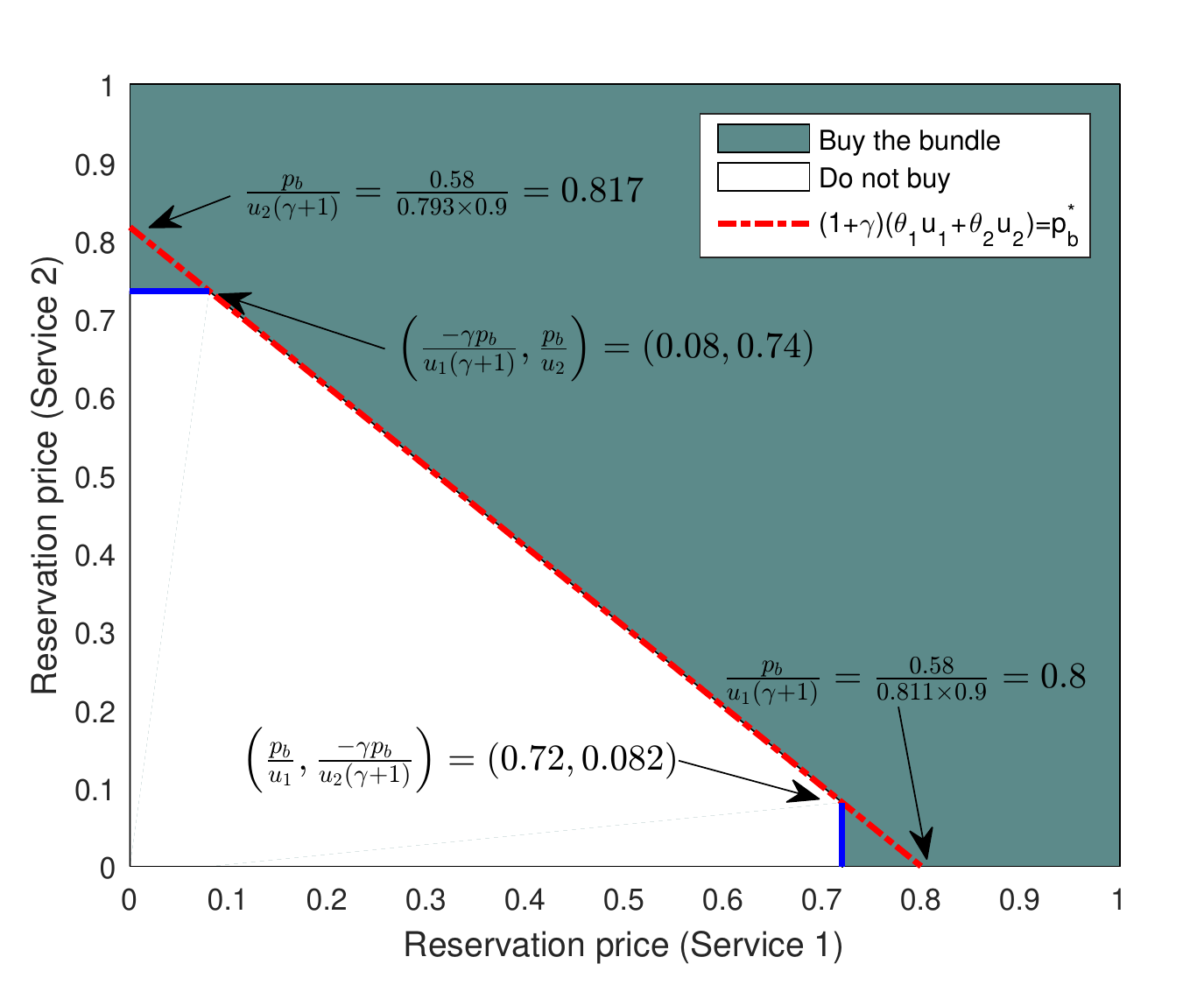}
\par\end{centering}
\caption{Demand boundaries of $S_{b2}$.\label{fig:substitutes_decision}}
\end{figure}

\subsubsection{Demand Boundaries}

Figure~\ref{fig:substitutes_decision} presents the demand on $S_{b2}$ consisting of substitute services. There are three decision boundaries. Firstly, the customers buy the bundle if their valuations lie above and to the right of the decision line $(1+\gamma)(\theta_{1}u_{1}+\theta_{2}u_{2})=p_{b}^{*}$, where $u_{1}=0.811$, $u_{2}=0.793$, $\gamma=-0.1$, and $p_{b}^{*}=0.58$. Secondly, the customers buy $S_{b2}$ if their valuation $\theta_{1}$ of $S_{1}$ is greater than or equal to $\frac{p_{b}}{u_{1}}$. Thirdly, the customers buy $S_{b2}$ if their valuations $\theta_{2}$ of $S_{2}$ are greater than or equal to $\frac{p_{b}}{u_{2}}$.

\begin{figure}
\begin{centering}
\includegraphics[width=1\columnwidth]{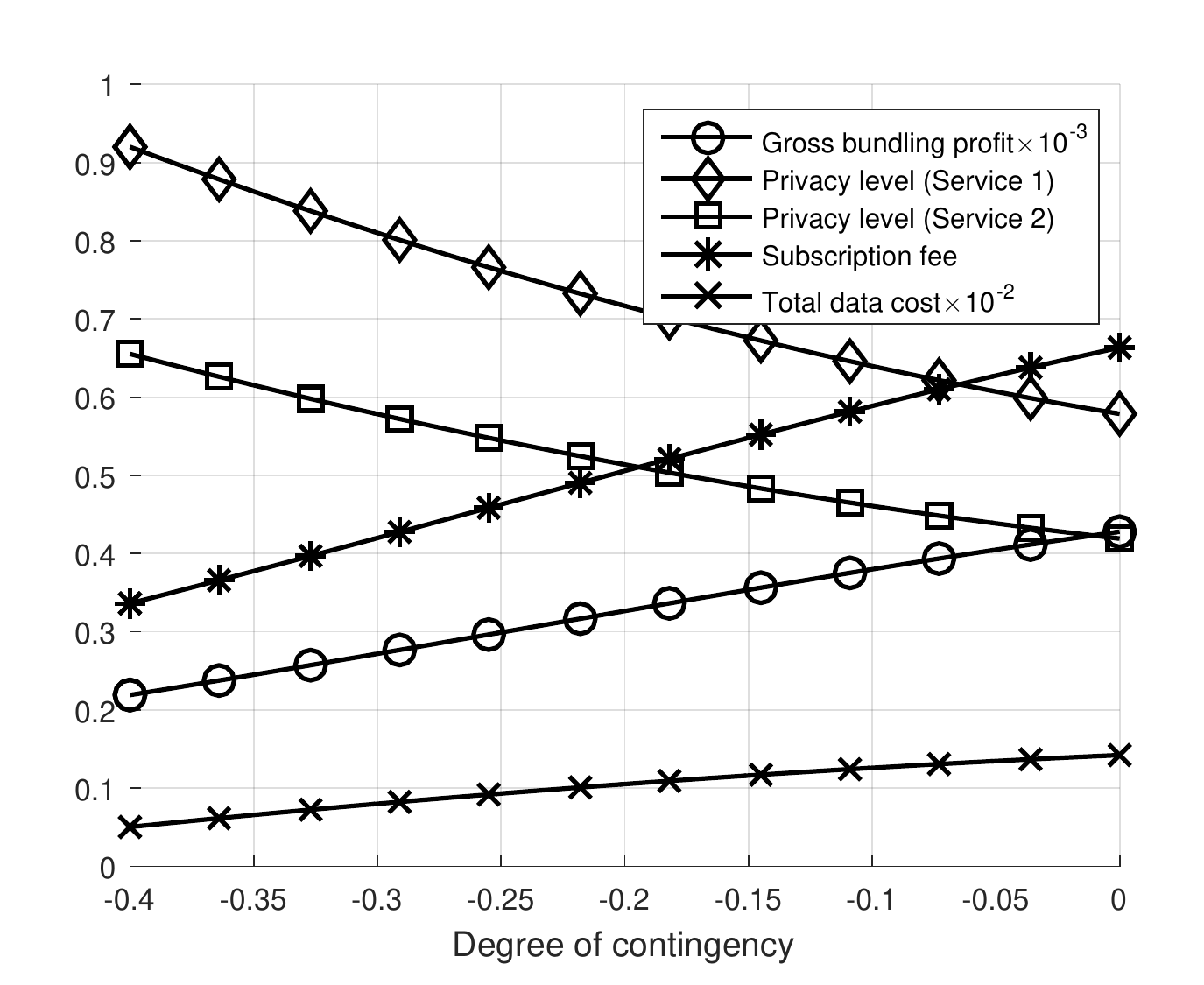}
\par\end{centering}
\caption{Impacts of the degree of contingency~$\gamma$ on the gross profit $G_{s}^{*}(\cdot)$, privacy levels~$r_{1}^{*}$ and $r_{2}^{*}$, subscription fee~$p_{b}^{*}$, and total data cost.\label{fig:substitutes_contingency}} 
\end{figure}

\begin{figure}
\begin{centering}
\includegraphics[width=1\columnwidth]{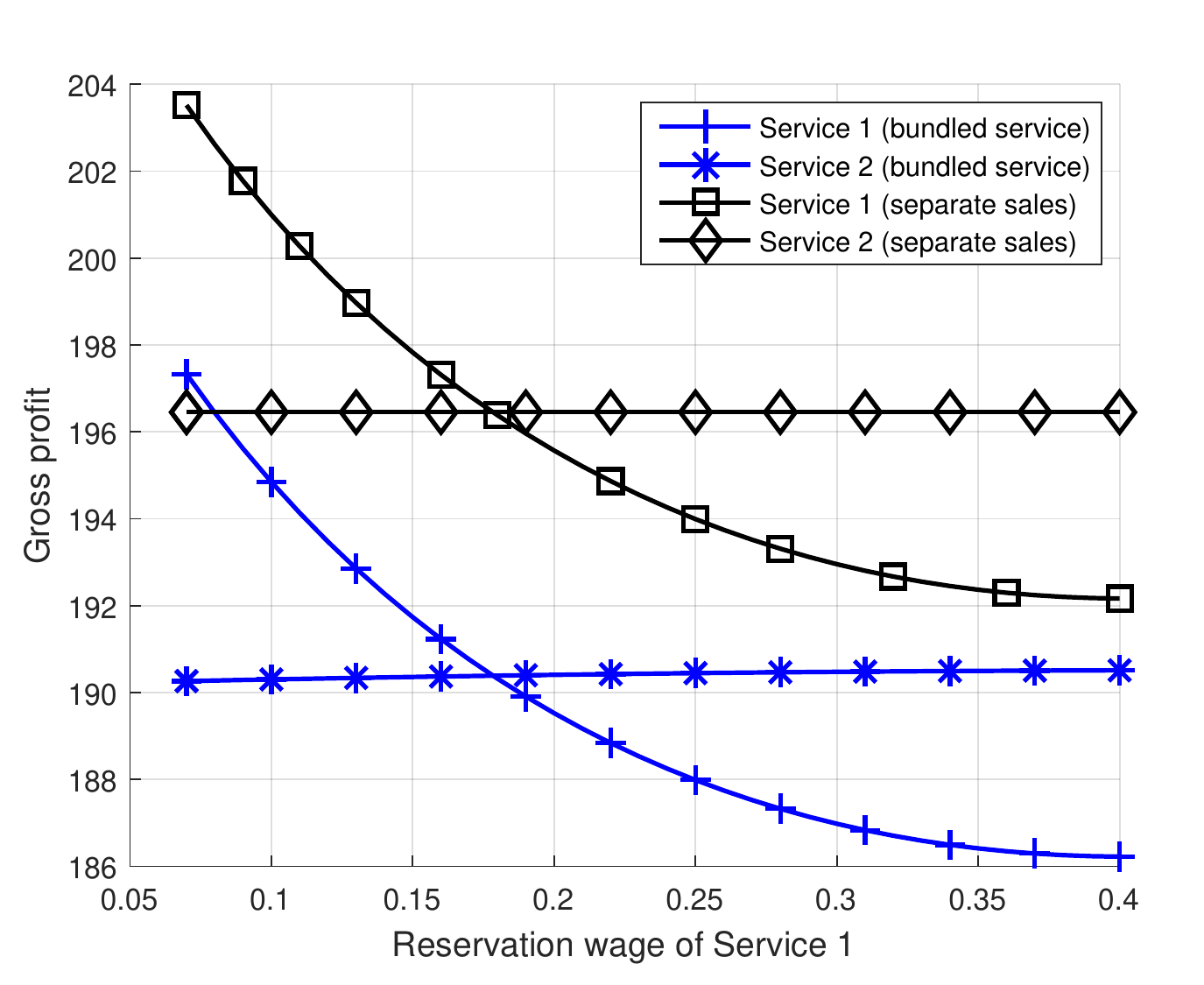}
\par\end{centering}
\caption{Profits of $S_{1}$ and $S_{2}$ when sold separately or bundled as substitutes.\label{fig:substitutes_profit_sharing}}
\end{figure}

\subsubsection{The Impact of Contingency Degree}

Interrelated products are modeled as substitutes when $\gamma<0$. Figure~\ref{fig:substitutes_contingency} shows that when the degree of contingency is decreased, the gross profit $G_{s}^{*}(\cdot)$, subscription fee~$p_{b}^{*}$, and total data cost decrease. This correlation is expected as decreasing $\gamma$ indicates high similarity among $S_{1}$ and $S_{2}$. Thus, the customer valuations of the resulting bundle decrease and the subscription fee moves to lower values accordingly.

\subsubsection{Profit Sharing}

Bundling substitute services is detrimental for the gross profit compared to the separate sales as shown in Figure~\ref{fig:substitutes_profit_sharing}. In particular, the customer valuation of $S_{b2}$ containing similar and comparable services is reasonably low. The bundling profit, therefore, falls below the total profits under the separate sales of $S_{1}$ and $S_{2}$ $G_{s}(r_{1}^{*},r_{2}^{*},p_{b}^{*})<F_{1}(r^{*},p_{s}^{*})+F_{2}(r^{*},p_{s}^{*})$. The rational service providers will decide to sell $S_{1}$ and $S_{2}$ separately.

\section{Conclusion and Future Work\label{sec:conclusions}}

In this paper, we have presented the profit maximization and pricing models for selling people-centric services separately and as service bundles. We have firstly modeled the tradoff between the service quality and privacy level from data analytics perspectives. Specifically, the service quality has been shown to be inversely proportional to the privacy level. For separately offered services, the service provider jointly optimizes the privacy level and subscription fee to maximize the gross profit resulting from offering services to a set of customers. For service bundling, the people-centric services are bundled as complementary and substitute services. Accordingly, the privacy levels of the two services and subscription fee are optimized to gain the maximum gross profit for the service provider. Finally, we have presented a model for sharing the resulting bundling profit among the cooperative people-centric services.

For the future work, the heterogeneity of the crowdsensing participants and competitive markets can be included in the profit maximization models.

\section*{Acknowledgment}

This work was supported in part by Singapore MOE Tier~1 (RG18/13 and RG33/12) and MOE Tier 2 (MOE2014-T2-2-015 ARC4/15 and MOE2013-T2-2-070 ARC16/14).  It was also supported in part by the U.S. National Science Foundation under Grants US NSF CNS-1646607, ECCS-1547201, CCF-1456921, CNS-1443917, and ECCS-1405121.

\bibliographystyle{IEEEtran}
\bibliography{ref}

\section*{Biographies}
\begin{IEEEbiographynophoto}
{Mohammad Abu Alsheikh}
[S'14] (mohammad027@e.ntu.edu.sg) received the B.Eng. degree in computer systems engineering from Birzeit University, Palestine, in 2011. Between 2010 and 2012, he was a Software Engineer working on developing robust web services, Ajax-based web components, and smartphone applications. He is currently a Ph.D. candidate in the School of Computer Science and Engineering, Nanyang Technological University, Singapore. His research interests include machine learning in big data analytics, mobile sensing technologies, and sensor-based activity recognition.
\end{IEEEbiographynophoto}
\vfill

\begin{IEEEbiographynophoto}
{Dusit Niyato}
[M'09--SM'15--F'17] (dniyato@ntu.edu.sg) is currently an Associate Professor in the School of Computer Science and Engineering, at Nanyang Technological University, Singapore. He received B.Eng. from King Mongkuts Institute of Technology Ladkrabang (KMITL), Thailand in 1999 and Ph.D. in Electrical and Computer Engineering from the University of Manitoba, Canada in 2008. His research interests are in the area of energy harvesting for wireless communication, Internet of Things (IoT) and sensor networks.
\end{IEEEbiographynophoto}

\begin{IEEEbiographynophoto}
{Derek Leong} (dleong@i2r.a-star.edu.sg) received the B.S. degree in electrical and computer engineering from Carnegie Mellon University in 2005, and the M.S. and Ph.D. degrees in electrical engineering from the California Institute of Technology in 2008 and 2013, respectively. He is a scientist with the Smart Energy and Environment cluster at the Institute for Infocomm Research~(I\textsuperscript{2}R), A*STAR, Singapore. His research and development interests include distributed systems, sensor networks, smart cities, and the Internet of Things.
\end{IEEEbiographynophoto}

\begin{IEEEbiographynophoto}
{Ping Wang}
[M'08--SM'15] (wangping@ntu.edu.sg) received the Ph.D. degree in electrical engineering from University of Waterloo, Canada, in 2008. Currently she is an Associate Professor in the School of Computer Science and Engineering, Nanyang Technological University, Singapore. Her current research interests include resource allocation in multimedia wireless networks, cloud computing, and smart grid. She was a corecipient of the Best Paper Award from IEEE Wireless Communications and Networking Conference~(WCNC) 2012 and IEEE International Conference on Communications~(ICC) 2007.
\end{IEEEbiographynophoto}

\begin{IEEEbiographynophoto}
{Zhu Han}
[S'01--M'04--SM'09--F'14] (zhan2@uh.edu) received the B.S. degree in electronic engineering from Tsinghua University, in 1997, and the M.S. and Ph.D. degrees in electrical engineering from the University of Maryland, College Park, in 1999 and 2003, respectively. From 2000 to 2002, he was an R\&D Engineer of JDSU, Germantown, Maryland. From 2003 to 2006, he was a Research Associate at the University of Maryland. From 2006 to 2008, he was an assistant professor in Boise State University, Idaho. Currently, he is a Professor in Electrical and Computer Engineering Department as well as Computer Science Department at the University of Houston, Texas. His research interests include wireless resource allocation and management, wireless communications and networking, game theory, wireless multimedia, security, and smart grid communication. Dr. Han received an NSF Career Award in 2010, the Fred W. Ellersick Prize of the IEEE Communication Society in 2011, the EURASIP Best Paper Award for the Journal on Advances in Signal Processing in 2015, several best paper awards in IEEE conferences, and is currently an IEEE Communications Society Distinguished Lecturer.
\end{IEEEbiographynophoto}
\vfill

\end{document}